\documentclass[aps,prd,onecolumn,superscriptaddress, nofootinbib, notitlepage]{revtex4-1}  
\usepackage{amssymb,amsmath,amsfonts,amsbsy}
\usepackage{color}
\usepackage{float}

\usepackage{enumerate}
\usepackage{graphicx}
\usepackage{pifont}

\usepackage{comment}
\usepackage{subfigure}
\usepackage{ulem}
\usepackage{cancel}
\usepackage{xcolor} 
\usepackage{aas_macros}

\RequirePackage[colorlinks,citecolor=red,urlcolor=red,linkcolor=red]{hyperref}
\usepackage{bm}

\def\d{\mathrm{d}}

\begin{document}

\title{Morphological Signatures of Gravitational Evolution, Redshift-Space Distortions, and Massive Neutrinos in Large-Scale Structure}

\author{Priya Goyal}
\affiliation{Korea Institute for Advanced Study (KIAS), 85 Hoegiro, Dongdaemun-gu, Seoul, Republic of Korea-02455}
\author{Stephen Appleby}
\affiliation{Asia Pacific Center for Theoretical Physics, Pohang 37673, Republic of Korea}
\affiliation{Department of Physics, POSTECH, Pohang 37673, Republic of Korea}
\affiliation{Korea Institute for Advanced Study (KIAS), 85 Hoegiro, Dongdaemun-gu, Seoul, Republic of Korea-02455}
\author{Pravabati Chingangbam}
\affiliation{Indian Institute of Astrophysics, Koramangala II Block,       
  Bangalore  560 034, India}
\affiliation{Korea Institute for Advanced Study (KIAS), 85 Hoegiro, Dongdaemun-gu, Seoul, Republic of Korea-02455}
\author{Changbom Park}
\affiliation{Korea Institute for Advanced Study (KIAS), 85 Hoegiro, Dongdaemun-gu, Seoul, Republic of Korea-02455}

\begin{abstract}
We investigate the morphological properties of large-scale structure in the Universe and the physical processes that modify the excursion-set morphology of the three-dimensional matter density field. Using the Quijote N-body simulation suite, we study how an initially Gaussian random matter density field is altered by non-linear gravitational evolution, redshift-space distortions, and massive neutrino free-streaming. To quantify these effects, we employ a comprehensive set of morphological descriptors, including Minkowski Functionals, Betti numbers, Minkowski Tensors, and local measures of the size and shape of connected components and cavities. We find that gravitational evolution, on quasi-linear scales $R_G \sim 10 h^{-1} \mathrm{Mpc}$, strongly skews the one-point distribution and slightly smooths the field via the merging of critical points, with a more pronounced effect for minima and wall saddle points than for peaks. Redshift-space distortions produce the strongest morphological signal, generating pronounced anisotropies that are robustly captured by Minkowski Tensors and local shape measures, arising from both coherent large-scale flows and non-linear Finger-of-God effects. In contrast, massive neutrinos induce an approximately isotropic suppression of small-scale structure, slightly reducing the amplitudes of the Minkowski Functionals while leaving individual shape measures largely unchanged. We further explore the sensitivity of these statistics to variations in cosmological parameters $\Omega_m$, $n_s$, and $\sigma_8$, finding that they probe strongly degenerate combinations of $\Omega_m$ and $n_s$, while also exhibiting sensitivity to $\sigma_8$ through the non-Gaussianity of the evolved density field.  
\end{abstract}

\maketitle

\section{Introduction}
\label{sec:introduction}

The large-scale structures (LSS) in the Universe contains a wealth of cosmological information. As primordial density fluctuations evolve under gravitational instability, they form the complex web of clusters, filaments, walls, and voids observed in the present-day matter distribution. The cosmological information is encoded in the statistical properties of the LSS, which is typically extracted using the power spectrum (and its Fourier counterpart, correlation function)~\citep{1980lssu.book.....P} and higher $N$-point functions. To probe the non-linear and non-Gaussian regime of structure formation, higher-order statistics~\citep{1980lssu.book.....P,2002PhR...367....1B} and geometrical descriptors~\citep{1994A&A...288..697M,Schmalzing:1995qn,JMI:JMI3331,Matsubara:2000dg,2013NJPh...15h3028S} are increasingly utilized.
One such approach is to study the morphology of excursion sets of the field, characterising them by their geometrical and topological properties.

Topological measures like the genus, which quantifies the connectivity of iso-density or iso-temperature contour surfaces of smooth matter density or the cosmic microwave temperature field, and the Betti numbers, which are the counts of individual connected components and cavities, have long been utilized to probe primordial non-gaussianity~\citep{1996ApJ...463..409M,Hikage:2006fe,PC_2009,Park:2013dga}, reconstruct expansion history of the universe~\citep{Blake2013UsingTT,Speare_2015} and constrain models of galaxy formation and cosmological parameters~\citep{MELOTT19901,2000ApJ...529..795C,GottIII_2008,Choi_2010,Appleby:2020pem}. Alongside topology, the geometry of excursion sets is often characterized using Minkowski functionals (MFs) which provide information on the statistical properties of the field and the physical processes that shape it~\citep{1994A&A...288..697M,Schmalzing:1995qn,https://doi.org/10.1111/j.1365-2966.2008.13674.x,2016MNRAS.461.1363N,Chingangbam:2017PhLB,Planck:2019evm,Appleby:2021xoz,Gott:1989yj,1991ApJ...378..457P,Schmalzing:1997aj,Schmalzing:1997uc,1989ApJ...345..618M,1992ApJ...387....1P,Kerscher:1998gs,2001ApJ...553...33P,Park:2009ja,doi:10.1111/j.1365-2966.2010.18015.x,Sahni:1998cr,Bharadwaj:1999jm,2003ApJ...584....1M,vandeWeygaert:2011ip, Park:2013dga,vandeWeygaert:2011hyr,Chingangbam:2013,Shivshankar:2015aza,Pranav:2016gwr,Pranav:2018lox,
Pranav:2018pnu,Feldbrugge:2019tal,Wilding:2020oza,Munshi:2020tzm,Liu:2022vtr,Liu:2023qrj,Liu:2025haj,Rana:2018,Rahman:2021,Kanafi:2023hmr,10.1093/mnras/staf1110}. In addition to MFs, the Minkowski Tensors (MTs) are their tensorial generalisation, extend the morphological analysis by capturing directional and anisotropic information of the field. MTs are defined as integrals over the boundary of an excursion set, with integrands proportional to symmetric tensor products of position vectors and normals to the boundary~\citep{Alesker1999,Beisbart:2001vb,Beisbart:2001gk,2002LNP...600..238B,HUG2008482,JMI:JMI3331,2013NJPh...15h3028S}. 
The ensemble average of the MTs measured from Gaussian, isotropic and anisotropic random fields, were derived in~\citep{Chingangbam:2017uqv,2019ApJ...887..128A,2019ApJ...887..128A,Chingangbam:2021}. Further, authors in~\citep{Klatt_2022} analytically calculated expected MTs of arbitrary rank for the level sets of Gaussian random fields. Numerical methods with which to extract the MTs from random fields can be found in~\citep{JMI:JMI3331,2018ApJ...858...87A,Schaller2020,Collischon:2024jhw}. These statistics were first applied to cosmological data
in~\citep{Ganesan:2017}, where a significant anisotropic signal was found in the 2015 E-mode Planck data. 
Subsequently, they were employed to investigate different cosmological fields including CMB temperature and polarization fields~\citep{Chingangbam:2017PhLB,K.:2018wpn,Joby:2021,Goyal:2019vkq,Goyal:2021nun}, 21cm brightness temperature field from reionization epochs~\citep{Kapahtia:2019ksk,Kapahtia:2021}, CMB foregrounds like synchrotron radiation and dust emission~\citep{Rahman:2021,2024JCAP...01..036R} among many. In addition, authors have written a series of papers on the application of the MTs to the low redshift matter density field as traced by galaxies~\citep{2018ApJ...863..200A,2018ApJ...858...87A,Appleby:2021xoz}. MTs are particularly useful as they provide directional information that is not present in the scalar Minkowski functionals.

In this paper, we aim to investigate how various cosmological effects: non-linear gravitational evolution, redshift-space distortion (RSD), and the presence of massive neutrinos, alter the morphology of the excursion sets of the three-dimensional matter density field relative to its initial, Gaussian state. Each of these phenomena introduces characteristic modifications to the matter density field. Gravitational evolution drives mode coupling and enhances non-Gaussianity, RSD induces anisotropy in the observed density field through velocity-induced distortions along the line of sight, and free-streaming massive neutrinos damps small-scale power. Each will imprint a characteristic signal on the morphology of excursion sets that we will quantify in what follows. We focus on scales $R_{G} \sim 10 \, h^{-1} \, {\rm Mpc}$ throughout, which are simultaneously weakly non-Gaussian while also being poorly described by standard perturbation theory. These are precisely the scales that can be probed by the current generation of galaxy surveys.  

A detailed analysis in~\citep{Kim_2014} precisely quantified the impact of various systematic effects on the genus statistic of gravitationally evolved matter and dark matter halo density fields. The study considered finite pixel size, non-linear gravitational evolution, redshift-space distortion (RSD), shot noise due to finite sampling, and halo bias. In a similar study, analysis of the two-dimensional genus statistic of the galaxy distribution using Horizon Run 4 simulations~\citep{Kim:2015yma} focused on similar systematic effects, finding that while pixel and gravitational effects are relatively minor $(\le 1\%)$, RSD and shot noise can induce genus amplitude changes up to $\sim10\%$, though these can be mitigated by thick redshift shells and mass-based galaxy sampling~\citep{2017ApJ...836...45A}. Several studies have also explored, in a quantitative framework, the potential of leveraging the morphological properties of large-scale structure (LSS) to constrain the sum of neutrino masses, $M_\nu$. These works utilize statistics such as Minkowski Functionals, power spectrum, bispectrum, marked statistics to capture the imprints of massive neutrinos (scale-dependent suppression of small-scale structure) in the morphology of matter density or galaxy density fields. They demonstrate that LSS morphology can serve as a complementary probe to traditional power spectrum–based methods in constraining $M_{\nu}$ and simultaneous measurement of both yields significant improvement over the power spectrum constraints~\citep{Marques:2018ctl,2020PhRvD.101f3515L,Liu:2022vtr,Liu:2023qrj}.

The present analysis is focused on characterizing the deviations in morphology and topology induced by the above mentioned physical phenomena, rather than constraining cosmological parameters. To quantify these effects, we employ a set of statistical measures including Minkowski Functionals, Betti numbers, and Minkowski Tensors. This work is a purely numerical study based on the Quijote simulations~\citep{Villaescusa-Navarro:2019bje}, a large-scale suite of N-body simulations designed for cosmological inference using machine learning and statistical methods. The availability of thousands of realizations with controlled variations of cosmological parameters, including the neutrino mass, allows us to isolate the impact of each physical effect on the morphology of the matter distribution. 

This paper is organized as follows. In Section~\ref{sec:data}, we provide a brief overview of the Quijote simulation data used in our analysis and introduce the set of morphological statistics employed—namely, Minkowski Functionals, Minkowski Tensors, Betti numbers, as well as the averaged properties of individual connected components and holes. Section~\ref{sec:grf} focuses on the morphological properties of Gaussian random fields (GRFs), which represent the initial conditions of the Universe at some unspecified initial time. The impact of gravitational collapse, redshift space distortion and massive neutrinos on our statistics are presented in Sections~\ref{sec:grav},~\ref{sec:rsd} and~\ref{sec:neutrinos}, respectively. We describe the response of our summary statistics to cosmological parameters in Section~\ref{sec:cosparams} and summarize our results in Section~\ref{sec:discuss}. 

\section{Data and Methodology} 
\label{sec:data}

We use dark matter simulations to study how various cosmological processes affect the morphology and topology of the matter density field. In this section we review the data used in our analysis and the summary statistics that we measure. 

\subsection{Data} 

We use the Quijote simulations for this study~\citep{Villaescusa-Navarro:2019bje}, which are an ensemble of cosmological scale dark matter simulations; $\sim 44,000$ realisations of $512^{3}$ particles gravitationally evolved in $L= 1000 \, h^{-1} {\rm Mpc}$ boxes from $z=127$ to $z=0$. Specifically, we use $N_{\rm real} = 50$ redshift zero snapshot boxes and construct real space density fields by binning the dark matter particles into a uniform $512^3$ Cartesian grid of resolution $\Delta = 2 \, h^{-1} \, {\rm Mpc}$. We then define the number density field $\delta_{\{i,j,k\}} = (n_{\{i,j,k\}} - \bar{n})/\bar{n}$, where $n_{\{i,j,k\}}$ is the number of particles in the $\{i,j,k\}$ pixel and $\bar{n}$ is the mean number of particles per pixel. Throughout this work, we refer to the particle number density field simply as the matter density field. We smooth this field in Fourier space using a Gaussian kernel $W(kR_{\rm G}) = e^{-k^{2}R_{G}^{2}/2}$. After smoothing, the field is again mean-subtracted and normalised to unit variance. We adopt the smoothing scale $R_{\rm G} =10 \, h^{-1} \, {\rm Mpc}$ throughout this work, which is the typical mean galaxy separation of current generation galaxy surveys. 

The Quijote simulations are performed using a wide range of different cosmological parameter sets. We focus predominantly on the fiducial, Planck $\Lambda$CDM cosmology with parameters $\Omega_{m} = 0.3175$, $n_{s} = 0.9624$, $\Omega_{b} = 0.049$, $h = 0.6711$, $\sigma_{8} = 0.834$, but also consider variations of cosmological parameters in Section~\ref{sec:cosparams} and the inclusion of massive neutrinos in Section~\ref{sec:neutrinos}. Details of how massive neutrinos were modelled in the Quijote simulations can be found in \cite{2017MNRAS.466.3244Z}. 

To compare our results to the initial condition of the matter density field, we generate $N_{\rm real} = 50$ realisations of a Gaussian random field with the same cosmological parameters as the fiducial Quijote simulations. The Gaussian fields are generated on a regular grid in Fourier space and inverse fast Fourier transformed using the same size and resolution as the Quijote snapshot boxes. They are finally smoothed with $R_{\rm G} =10 \, h^{-1} \, {\rm Mpc}$, mean-subtracted and unit variance normalized exactly as the Quijote data, and the same statistics extracted. 

The summary statistics are extracted from the fields using the numerical procedures described in \cite{2018ApJ...863..200A}, using a marching tetrahedron algorithm to generate iso-field surfaces. We direct the reader to Appendices A and B of that work for further details.

\subsection{Minkowski Functionals, Betti numbers, Minkowski Tensors}

Given a mean-subtracted, smoothed dark matter density field $\delta(\mathbf{x})$ with variance 
$\sigma_{0}^{2} = \langle \delta^2 \rangle$, we define the excursion set at threshold \(\nu\) as $Q(\nu) = \{ \mathbf{x} \, | \, \delta(\mathbf{x}) \geq \nu \sigma_0 \}$. 
The set $Q(\nu)$ contains all spatial regions where the density contrast exceeds a specified threshold. We denote its boundary by $\partial Q$. 
For a typical $\nu$, $Q$ is not one continuous region but rather consists of disjoint subsets or elements, $Q_{i}$, with non-intersecting boundaries $\partial Q_{i}$, where $i$ indexes each element. Throughout this work we call each individual element as a {\it connected component}, following the mathematics nomenclature\footnote{In the cosmology literature, the disjoint elements are also referred to as `excursion regions'.}. Similarly, we can construct excursion sets using all spatial points equal to or below the threshold, as, $\widetilde{Q}(\nu) = \{ \mathbf{x} \, | \, \delta(\mathbf{x}) \leq \nu \sigma_0 \}$.  Akin to ${Q}$, $\widetilde{Q}$ is composed of disjoint elements $\widetilde{Q}_{i}$, with non-intersecting boundaries $\partial \widetilde{Q}_i$. We refer to each $\widetilde{Q}_{i}$ as a {\it cavity}, or interchangeably {\it hole}, since $\widetilde{Q}_i$'s are manifested as cavities inside connected components~(see figure 2 of \cite{Chingangbam:2025gcl}). Both connected components and cavities can be either simply or multiply connected, and here we make no distinction regarding their topology.  Both $\partial Q_i$ and $\partial\widetilde{Q}_i$ are orientable surfaces, but they are distinguished by opposite sense of orientation.

By varying $\nu$, one can probe the morphology of a given field across different threshold  levels - from high-density peaks $(\nu \gg 0)$ to deep minima $(\nu \ll 0)$. In this paper, to characterize the geometric and topological properties of excursion sets,   
we employ two different families of morphological descriptors that can be classified as: (1) volume averaged statistics defined for the total excursion set $Q$, and, (2) local statistics that quantify the morphology of individual $Q_i$ and $\widetilde Q_i$. The definitions for these two classes of statistics are given below.

\subsubsection{Volume averaged statistics}

{\bf Minkowski functionals}: The most widely adopted statistics in cosmology that utilize excursion sets are the Minkowski Functionals (MFs), which encode the morphology and topology of a field. In three dimensions we define the four MFs as 
\begin{equation}
			W_0 = \frac{1}{V} \int_{Q} dV \, , \hspace{10mm}	W_1 = \frac{1}{6 V} \int_{\partial Q} dA \, , \hspace{10mm}    W_2 = \frac{1}{3 \pi V} \int_{\partial Q} G_2 dA \, , \hspace{10mm}    W_3 = \frac{1}{4 \pi^2 V} \int_{\partial Q} G_3 dA  \, ,
\end{equation}

\noindent where $\partial Q$ is the boundary $\delta = \nu\sigma_{0}$, $V$ is the total volume of the domain and $G_{2} = (R_{1}^{-1} + R_{2}^{-1})/2$ and $G_{3} = (R_{1}R_{2})^{-1}$, with $R_{1},R_{2}$ being the principal curvatures of the surface $\partial Q$. These statistics respectively encode the volume enclosed, surface area, mean and Gaussian curvatures of the excursion set boundary. They are defined here as densities, and as such can be related to local one-point cumulants of the field under the assumption of statistical homogeneity. 

{\bf Betti numbers}: Among the MFs, $W_3$ is a purely topological quantity that is invariant under continuous deformations of the excursion sets. It can be  expressed in terms of three Betti numbers, which further characterize the topology of these sets. Specifically, $b_{0}$ denotes the number of connected components $Q_i$ of an excursion set, and $b_{2}$ the number of distinct  cavities $\widetilde Q_i$. The third  one, $b_{1}$, denotes the number of tunnels or loops which measure the degree of multiply-connectedness of $Q_i$ and $\widetilde Q_i$. The alternating sum of the Betti numbers is then related to $W_3$ ~\citep{Park:2013dga} via
\begin{equation}
W_3 \propto -b_0+b_1 -b_2.
\end{equation}
We direct the reader to \citep{vandeWeygaert:2011hyr,Pranav:2018pnu,Feldbrugge:2019tal} for a detailed description of these statistics and other related topological measures such as persistence diagrams. 

{\bf Minkowski Tensors}: These are tensorial generalizations of MFs that capture directional information. We consider two translation invariant, rank-2 MTs defined as 
\begin{equation}
			W^{0,2}_1 = \frac{1}{6 V} \int_{\partial Q} \hat{n} \otimes \hat{n} dA \, , \hspace{10mm}
            W^{0,2}_2 = \frac{1}{3 \pi V} \int_{\partial Q} G_2 \hat{n} \otimes \hat{n} dA 
\end{equation}

\noindent where $\hat{n}$ is the unit normal to $\partial Q$ and $\otimes$ denotes the symmetric tensor product. For the purpose of the present study, these are $3\times 3$ Cartesian tensors, although see~\citep{Appleby:2022itn} for a discussion of the plane parallel limit and radial redshift space distortions. For the coordinate system chosen in this work (Cartesian basis aligned with snapshot box axes), all information is contained in the diagonal elements of these matrices, which we denote as $W^{0,2}_{1}|_{\mu}{}^{\mu}$ and $W^{0,2}_{2}|_{\mu}{}^{\mu}$ with $\mu=1,2,3$.  In redshift space, rotating the coordinate system into misalignment with the line-of-sight would generate off-diagonal elements, but for real space, isotropic fields, both tensors are proportional to the identity matrix. We direct the reader to \cite{2013NJPh...15h3028S,2018ApJ...863..200A} for a detailed definition of these quantities and associated numerical methodologies in three dimensions. 

\subsubsection{Morphology of individual connected components and cavities}

We next focus on the morphology of individual connected components $Q_i$ and cavities $\widetilde Q_i$, and define three summary statistics that quantify their average morphology.
The first is the effective radius of connected components and cavities, which are given by : 
\begin{eqnarray} 
R_{\rm eff}^{ {\rm con}} &=& {1 \over b_{0}}\left( {3 \over 4\pi}\right)^{1/3} \sum_{i=1}^{b_{0}} \left(\int_{Q_{i}}dV\right)^{1/3}, \\
R_{\rm eff}^{{\rm cav}} &=& {1 \over b_{2}}\left( {3 \over 4\pi}\right)^{1/3} \sum_{i=1}^{b_{2}} \left(\int_{\widetilde{Q}_{i}}dV\right)^{1/3}  .
\end{eqnarray} 
These two statistics measure the typical sizes of individual elements by averaging over the sizes of all $Q_i$ and $\widetilde Q_i$, respectively,  as functions of threshold. An important property of these definitions is that the non-linear mapping
\begin{equation}
    \int_{Q_{i}}dV \mapsto \left(\int_{Q_{i}}dV \right)^{1/3} \, , 
\end{equation}
does not commute with the sum over elements. This means that the random variables $R_{\rm eff}^{{\rm con, \, cav}}$ cannot be expressed purely in terms of the volume averaged quantities $W_{0}$, $b_{0}$ and $b_{2}$. We elaborate on the information content of these statistics in Appendix~\ref{sec:app_c}. 

The remaining two statistics quantify the average shapes of individual $Q_i$ and $\widetilde Q_i$. For connected components, we first compute Minkowski tensors  for each  $Q_{i}$, 
\begin{equation}
			W^{0,2}_{1, \, i} = \frac{1}{6} \int_{\partial Q_{i}} \hat{n} \otimes \hat{n} dA \, , \hspace{10mm}
            W^{0,2}_{2, \, i} = \frac{1}{3 \pi} \int_{\partial Q_{i}} G_2 \hat{n} \otimes \hat{n} dA \, ,
\end{equation}
and their eigenvalues 
\begin{equation} \Lambda_{\mu, i}^{{\rm con} \, (1)} \, , \qquad \Lambda_{\mu, i}^{{\rm con} \, (2)} \, , \quad \mu = 1,2,3 \ ,  
\end{equation} 
following the ordering $\Lambda_{1,i} \leq \Lambda_{2,i} \leq \Lambda_{3,i}$ such that $\Lambda_{3,i}$ corresponds to the longest principal axis. Dimensionless shape parameters are then defined as \cite{Ganesan:2017} 
\begin{equation} \beta_{J, i}^{{\rm con} \, (1)} = {\Lambda_{J,i}^{{\rm con} \, (1)} \over \Lambda_{3,i}^{{\rm con} \, (1)}} \leq 1  \, , \qquad \beta_{J, i}^{{\rm con} \, (2)} = {\Lambda_{J,i}^{{\rm con} \, (2)} \over \Lambda_{3,i}^{{\rm con} \, (2)}} \leq 1 \, , \quad J = 1,2 
\end{equation} 
These ratios are scale-invariant and measure only shape anisotropy. Spherical objects have $\beta = 1$, whereas small $\beta$ indicates elongated structures. 

Averaging over the shape parameters of all connected regions gives the summary statistics
\begin{equation}
 \beta_{J}^{{\rm con} (1)} = {1 \over b_{0}} \sum_{i=1}^{b_{0}} \beta_{J, i}^{{\rm con} \, (1)}, \quad \beta_{J}^{{\rm con} \, (2)} = {1 \over b_{0}} \sum_{i=1}^{b_{0}} \beta_{J, i}^{{\rm con} \, (2)} .
\end{equation}
For cavities, we can perform exactly the same calculation, arriving at the corresponding average shape parameters, 
\begin{equation}
\beta_{J}^{{\rm cav} (1)} = {1 \over b_{2}} \sum_{i=1}^{b_{2}} \beta_{J, i}^{{\rm cav} \, (1)} , \quad \beta_{J}^{{\rm cav} \, (2)} = {1 \over b_{2}} \sum_{i=1}^{b_{2}} \beta_{J, i}^{{\rm cav} \, (2)} \, .
\end{equation} 
Similarly to the quantities $R_{\rm eff}^{{\rm con}, {\rm cav}}$, shape parameters $\beta_{J}^{{\rm con,cav} \, (1, 2)}$ are constructed by applying a non-linear transformation (tensor components $\to$ eigenvalues $\to$ ratios) to individual objects prior to averaging, and so cannot be expressed purely in terms of the volume-averaged statistics $W_{1}^{0,2}$, $W_{2}^{0,2}$, $b_{0}$, $b_{2}$. 

Geometrically, the tensors $W_{1}^{0,2}$ and $W_{2}^{0,2}$ measure the overall anisotropy of all excursion sets in a globally defined coordinate system. In contrast, the shape parameters $\beta_{J}^{{\rm con,cav} \, (1, 2)}$  are defined in a local coordinate system aligned with each object's principal axis, and the average is then performed over the corresponding number of objects. Thus, they provides a local measure of anisotropy of the field. 
In summary, we use the statistics $\left[W_{0}, W_{1},W_{2},W_{3}, b_{0}, b_{2}, W_{1}^{0,2},W_{2}^{0,2}\right]$ to characterise the global morphology of the density field, and $\left[  R_{\rm eff}^{{\rm con}, {\rm cav}},  \beta_{J}^{{\rm con,cav} \, (1, 2)}\right]$ 
to quantify the properties of individual connected components and cavities. 

\section{ Morphology  of Excursion Sets for GRF}
\label{sec:grf}

We begin with  the case of isotropic Gaussian random fields (GRFs) drawn from a  
$\Lambda$CDM power spectrum with the same cosmological parameters as the fiducial Quijote simulations. The morphological properties of these fields serve as a reference baseline, enabling us to isolate and quantify subsequent morphological deviations induced by cosmological processes such as gravitational collapse, redshift-space distortions, and the presence of massive neutrinos. All the statistics considered in this section are well studied in the literature, with the exception of the effective radius. 
For GRFs, analytic expressions for ensemble expectations of  MFs~\citep{10.1143/PTP.76.952} are given by,
\begin{eqnarray}
    W_k(\nu) &=& \frac{1}{(2\pi)^\frac{k+1}{2}} \frac{\omega_3}{\omega_{3-k}\omega_k} \left(\frac{\sigma_1}{\sqrt{3}\sigma_0}\right)^k e^{-\nu^2/2} H_{k-1}(\nu),
    \label{eq:gmfs}
\end{eqnarray}
where $k=0,1,2,3$, $\sigma_{1}^{2} = \langle |\nabla \delta|^{2} \rangle$, $H_0, H_1,H_2$ are probabilist Hermite polynomials, and $H_{-1}=\sqrt{\frac{\pi}{2}}e^{\frac{\nu^2}{2}}{\rm erfc}\left(\frac{\nu}{\sqrt{2}}\right)$. The $\omega_k$ constants are  
$\omega_0=1$, $\omega_1=2$, $\omega_2=\pi$ and $\omega_3 = \frac{4\pi}{3}$. The corresponding functional forms for global MTs are proportional to the MFs. Apart from the MFs and MTs, closed-form expressions are not known for the other statistics measured in this work.

\begin{figure}
    \centering
  \fbox{\large Gaussian Random Fields}\\
  \vspace{.2cm}
    \includegraphics[width=1\textwidth]{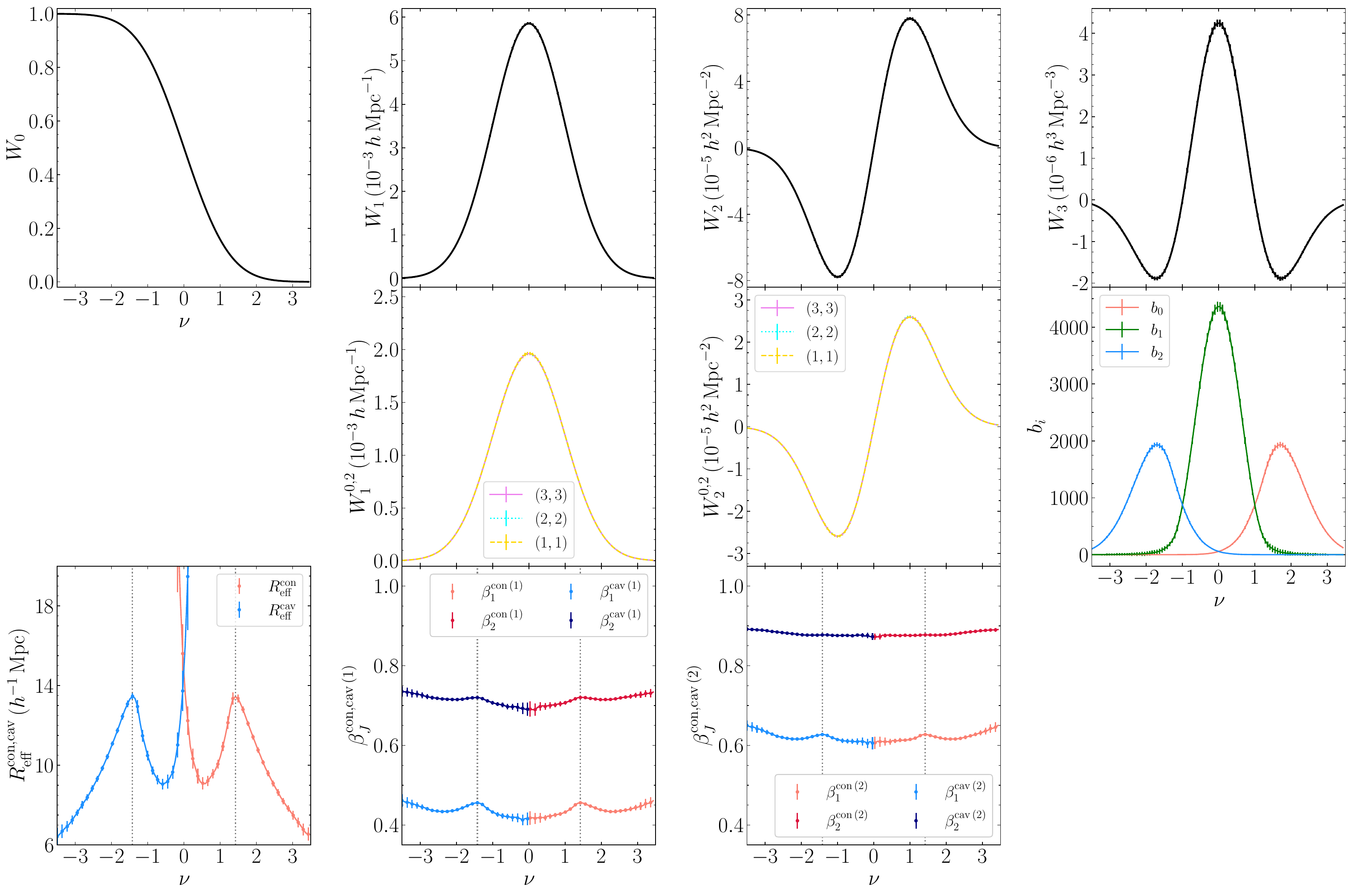} \\
\caption{Morphological statistics measured from Gaussian random field realisations smoothed on a scale of $R_G = 10\,h^{-1}\,\mathrm{Mpc}$ with a Gaussian filter.
\textbf{Top row:} The four Minkowski Functionals, shown by black curves with error bars indicating the $1\sigma$ scatter among realisations.
\textbf{Middle row:} The two Minkowski Tensors $W^{0,2}_1$ (left) and $W^{0,2}_2$(middle), and the three Betti numbers (right).
\textbf{Bottom row (left to right):} Local statistics: the effective size $(R^{{\rm con},{\rm cav}}_{\rm eff})$, and the shape parameters $(\beta_J^{{\rm con},{\rm cav}\,(1)})$ and $(\beta_J^{{\rm con},{\rm cav}\,(2)})$. Connected components and cavities are shown in red and blue, respectively, while darker and lighter curves correspond to the two $\beta$ eigenvalue ratios extracted from each Minkowski Tensor. }
    \label{fig:1_global}
\end{figure}

In the top row of Figure~\ref{fig:1_global}, we present the ensemble averaged Minkowski Functional curves $W_{0,1,2,3}$ of excursion sets of the GRFs, together with their $1\sigma$ uncertainties. These curves follow the expected functional form  given by eq. \ref{eq:gmfs}.
Due to the  $\delta \to - \delta$  invariance of the probability distribution of GRFs, the MFs exhibit  
the symmetries: $W_{0}(\nu) = 1 - W_{0}(-\nu)$, $W_{1}(\nu) = W_{1}(-\nu)$, $W_{2}(\nu) = - W_{2}(-\nu)$, $W_{3}(\nu) =  W_{3}(-\nu)$. Furthermore, all cosmological information is encoded in the amplitude of these curves, which are only sensitive to the ratio $r_{c} = \sigma_{0}/\sigma_{1}$. 

In the middle row of Figure~\ref{fig:1_global}, we show the diagonal elements of the Minkowski Tensors $W_{1}^{0,2}$ (left panel), $W_{2}^{0,2}$ (middle panel), together with the three Betti numbers $b_{0,1,2}$ (right panel). The diagonal components of each MT are shown as solid cyan/yellow/pink lines, which overlap exactly in this case, while the off-diagonal elements (not shown) are consistent with zero. This behaviour reflects the statistical isotropy of the Gaussian field, for which $W_{1}^{0,2} \propto W_{1}{\mathbb I}$, $W_{2}^{0,2} \propto W_{2}{\mathbb I}$, where ${\mathbb I}$ is the identity matrix. This is in agreement with the results reported in~\citep{Chingangbam:2021}, where the authors derive an analytic expression for the ensemble expectation of $W_1^{0,2}$ for two-dimensional fields. They show that for an isotropic field, $W_{1}^{0,2} \propto W_1 {\mathbb I}$, which indicates that there is no additional information in MTs compared to MFs in the absence of any preferred direction. 

The Betti numbers (second row, right panel) contain information that is not captured by the Minkowski Functionals \citep{MELOTT19901,10.1111/j.1365-2966.2011.18395.x, Chingangbam:2012,Park:2013dga, Pranav:2018pnu, Feldbrugge:2019tal, Wilding:2020oza, Pranav:2021ozq,Chingangbam:2025gcl}. At high thresholds, $\nu \gtrsim 2$, the genus is well approximated by $W_{3} \simeq -b_{0}/V$ while for low thresholds, $\nu \lesssim -2$, one finds $W_{3} \simeq -b_{2}/V$. In these regimes the topology of the excursion sets is relatively simple, being dominated by small isolated high-density peaks or low-density minima connected by saddle points. At intermediate thresholds $-2 \lesssim \nu \lesssim 2$, the topology becomes significantly more complex due to the presence of loops and embedded holes within larger connected regions. In this regime, the shape of the $b_{0,1,2}$ curves contain new information beyond the alternating sum $W_{3} \sim (-b_{0} + b_{1} - b_{2})/V$. Quantifying this statement is difficult, because closed-form expressions for ensemble averages of $ b_{0,1,2}$ are not known. 
However, near $\nu \approx 0$, one has $W_3 \approx b_{1}/V$. 
Because each Betti number dominates at different threshold levels, \cite{1992ApJ...392L..51P,2005ApJ...633....1P} introduced genus-related statistics—such as the cluster and void abundance parameters $A_C$ and $A_V$, and the shift parameter $\Delta\nu$—which summarize the information contained in $b_0$, $b_2$, and $b_{1}(\nu)$, respectively~\citep{Park:2013dga}.

Unlike $W_{3}$, the Betti numbers cannot be written exactly in terms of local one-point cumulants of the field and its derivatives for any threshold, as they quantify the connectivity of critical points that lie above a given threshold. A previous study of Gaussian random fields has revealed that the shape of the Betti curves $b_{0,1,2}(\nu)$ is sensitive to the shape of the underlying power spectrum of the field \citep{Park:2013dga}. This is in contrast to $W_{3}(\nu)$, for which the shape is exactly fixed by $W_{3}(\nu) \propto (1-\nu^{2})\exp[-\nu^{2}/2]$, and only the amplitude is sensitive to the power spectrum.  

In the bottom panels of Figure~\ref{fig:1_global}, we show effective radii $R_{\rm eff}^{{\rm con},{\rm cav}}$ (left), and the shape parameters $\beta_{J}^{{\rm con},{\rm cav} (1)}$ (middle) and $\beta_{J}^{{\rm con},{\rm cav} (2)}$ (right). All three statistics exhibit distinctive maxima and minima as functions of $\nu$, which reflect the changing morphology of the excursion sets. To describe this behaviour, we focus on $R_{\rm eff}^{{\rm con}}$ for the connected components (cf. red curve, bottom left panel). The interpretation for the cavities (blue curve) follows analogously under $\nu \to -\nu$. 

At large thresholds $\nu \gtrsim 3$, the excursion set is dominated by  isolated high-density peaks. As $\nu$ is lowered, two competing effects influence $R_{\rm eff}^{{\rm con}}$. First, existing objects grow in volume and merge as saddle points enter the excursion set, thereby increasing $R_{\rm eff}^{{\rm con}}$. Second, new and initially small isolated peaks appear, which has the effect of decreasing $R_{\rm eff}^{{\rm con}}$. 
In the range $\nu \gtrsim 2$, the first effect dominates, because the number of saddle points entering exceeds the number of new peaks (see the bottom left panel of Fig. 2 of ~\cite{Pogosyan:2009}), leading to an overall increase of $R_{\rm eff}^{{\rm con}}$ with decreasing $\nu$. 

At $\nu \simeq 1.4$, $R_{\rm eff}^{{\rm con}}$ exhibits a distinct maximum. At this threshold, the largest excursion subsets amalgamate into a single structure, removing most of the large objects from the average and replacing them with a single contribution. Because $R_{\rm eff}^{{\rm con}}$ averages over the fractional power $V_{i}^{1/3}$, larger objects are effectively de-weighted and therefore do not dominate the average. This amalgamation reduces the mean while the more numerous small elements remain distinct. The effective radius therefore decreases until $\nu \sim 0.7$. Below this value, the percolating structure rapidly grows and dominates the excursion set, linking nearly all components. At this point, there is a second rapid turnaround in $R_{\rm eff}^{{\rm con}}$ and the statistic approaches the box size.

The $\beta_{J}^{{\rm con},{\rm cav} \, (1,2)}$ curves (bottom row, middle and right panels) encode the same morphological information as functions of threshold, but in terms of shape rather than size. The overall variation of the curves with $\nu$ is weaker than that of $R_{\rm eff}^{{\rm con},{\rm cav}}$. For example, in the middle panel we have $\beta_{1}^{{\rm con} \, (1)} \sim \beta_{1}^{{\rm cav} \, (1)} \sim 0.45$ and $\beta_{2}^{{\rm con} \, (1)} \sim \beta_{2}^{{\rm cav} \, (1)} \sim 0.75$, with only a $\sim 10\%$ variation across the full range of $\nu$ values considered. This indicates that the average connected components and cavities are well described as triaxial ellipsoids with nearly constant axis ratios. The increasing number of saddle points entering the excursion set causes $\beta_{J}^{{\rm con} \, (1,2)}$ to decrease slightly from high to intermediate thresholds, and the extrema at $\nu \simeq \pm 1.4$ coincide with the peak in $R_{\rm eff}^{{\rm con},{\rm cav}}$, arising from the same merging and percolating behaviour of the field. At low thresholds $|\nu| \lesssim 1.4$, the shape functions $\beta_{J}^{{\rm con},{\rm cav} \, (1,2)}$ are even less sensitive to the presence of the large percolating excursion set because the averages are performed over eigenvalue ratios, meaning that objects of all sizes are uniformly weighted in the average. For more information on the properties of the individual connected components and cavities, we direct the reader to Appendices \ref{sec:app_c} and \ref{sec:app_th}.

In summary, the Minkowski Functionals (cf. top panels) probe the joint PDF of the field and its first and second derivatives, while the Minkowski tensors (second row, first two panels) probe the directional dependence. The Betti numbers (second row, right panel) contain information at threshold ranges $-2 \lesssim \nu \lesssim 2$ that is not fully captured by MFs and MTs. The average size and shape properties of connected components and cavities (bottom panels) characterize the percolation of the excursion sets as a function of threshold, providing a complementary perspective on the connectivity and large-scale organization of the field. 

\section{Gravitational evolution} 
\label{sec:grav}

In this section, we explore the impact of gravitational evolution on the morphology of excursion sets by comparing the non-linear matter density field at redshift $z = 0$ with the corresponding GRFs. We smooth the field on the relatively large scale $R_{G} = 10 \, h^{-1} \, {\rm Mpc}$. Note that at this scale we are not probing non-linear structures such as clusters or voids as conventionally defined in the literature. Those objects are defined via at least one shell crossing event, indicating collapse along one or more dimensions. Our large-scale, isotropic smoothing kernel ensures that we remain in the single-stream, weakly non-Gaussian regime. At the same time, standard perturbation theory and the Edgeworth expansion of the field PDF are known to perform poorly for $R_{G} \lesssim 10 \, h^{-1} \, {\rm Mpc}$ (see e.g. Figure 4 of \cite{Matsubara:2020knr}). Our choice of $R_{G}$ is deliberately selected to probe an intermediate regime: below the scales that can be accurately described by perturbation theory, yet sufficiently large that the field remains single-stream. 

\begin{figure}[h!]
    \centering
      \fbox{\large Gravitationally Evolved Dark Matter, $\nu$-threshold}\\
       \vspace{.2cm}
    \includegraphics[width=0.98\textwidth]{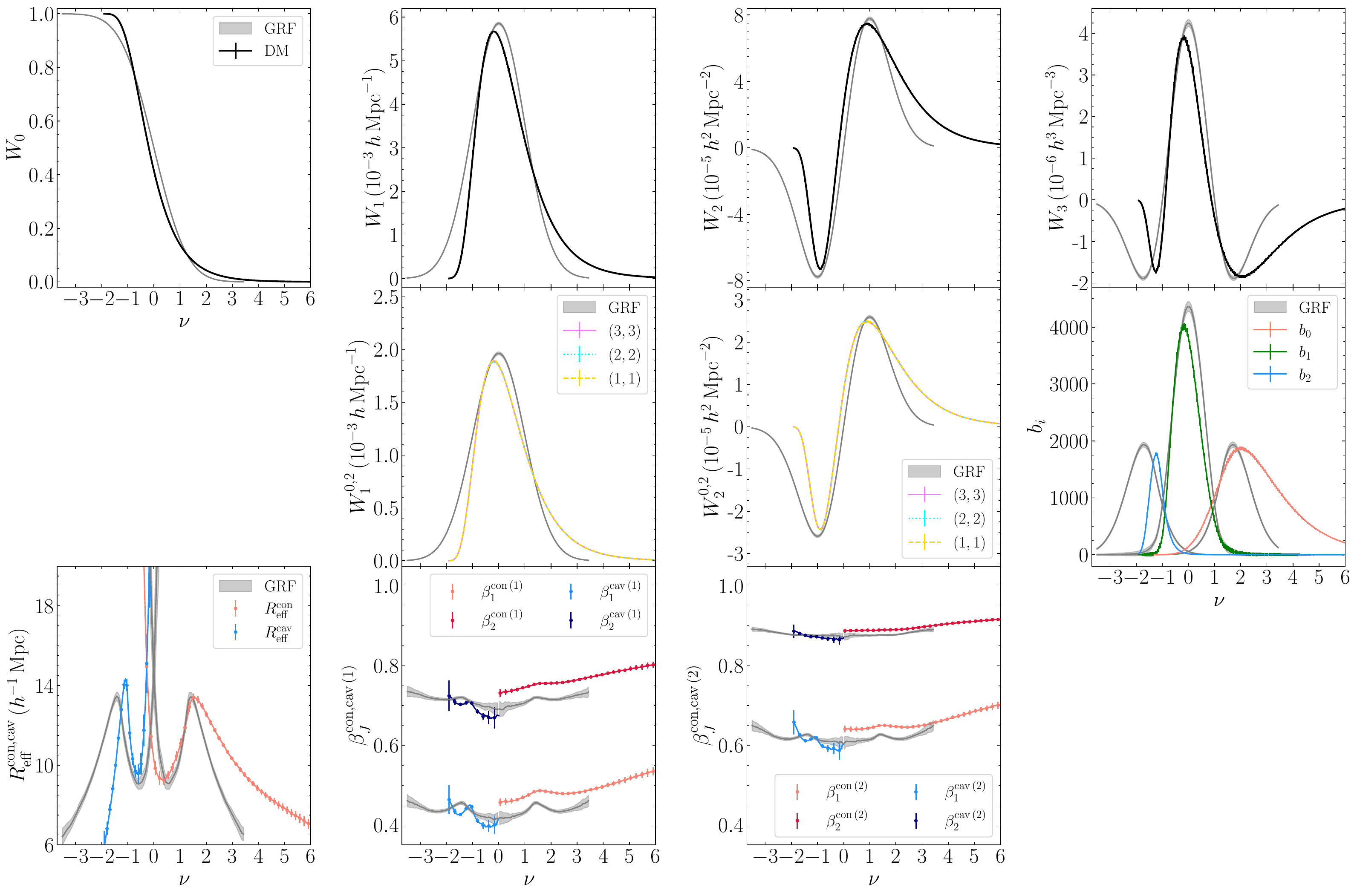}
    \caption{ Morphological statistics measured for gravitationally evolved dark matter fields as a function of iso-field threshold $\nu$. Coloured curves show the dark matter results, while the grey curves correspond to the Gaussian case from Figure~\ref{fig:1_global} and are included for comparison. The panel layout and colour scheme follow Figure~\ref{fig:1_global}. } 
    \label{fig:2_global}
\end{figure}

The first point to note is that the overwhelmingly dominant effect of gravitational evolution on the one-point statistics of the density field is to skew the PDF towards high positive values. This arises due to odd cumulants $\langle \delta^{n}\rangle$, starting with the bispectrum $n=3$, being induced via non-linear mode coupling. To illustrate this effect, in Figure~\ref{fig:2_global}, we present the same statistics as in the previous section, now for the $z=0$ dark matter snapshots. The colour scheme for all panels is the same as in Figure~\ref{fig:1_global}. For comparison, in every panel we also exhibit the Gaussian random field measurements from the previous section as grey filled bands. The positive skewness of the evolved field is clearly the most dominant feature of the coloured curves. 

 \begin{figure}[h!]
    \centering
    \fbox{\large Gravitationally Evolved Dark Matter, $\nu_{v}$-threshold}\\
    \vspace{.2cm}
    \includegraphics[width=0.98\textwidth]{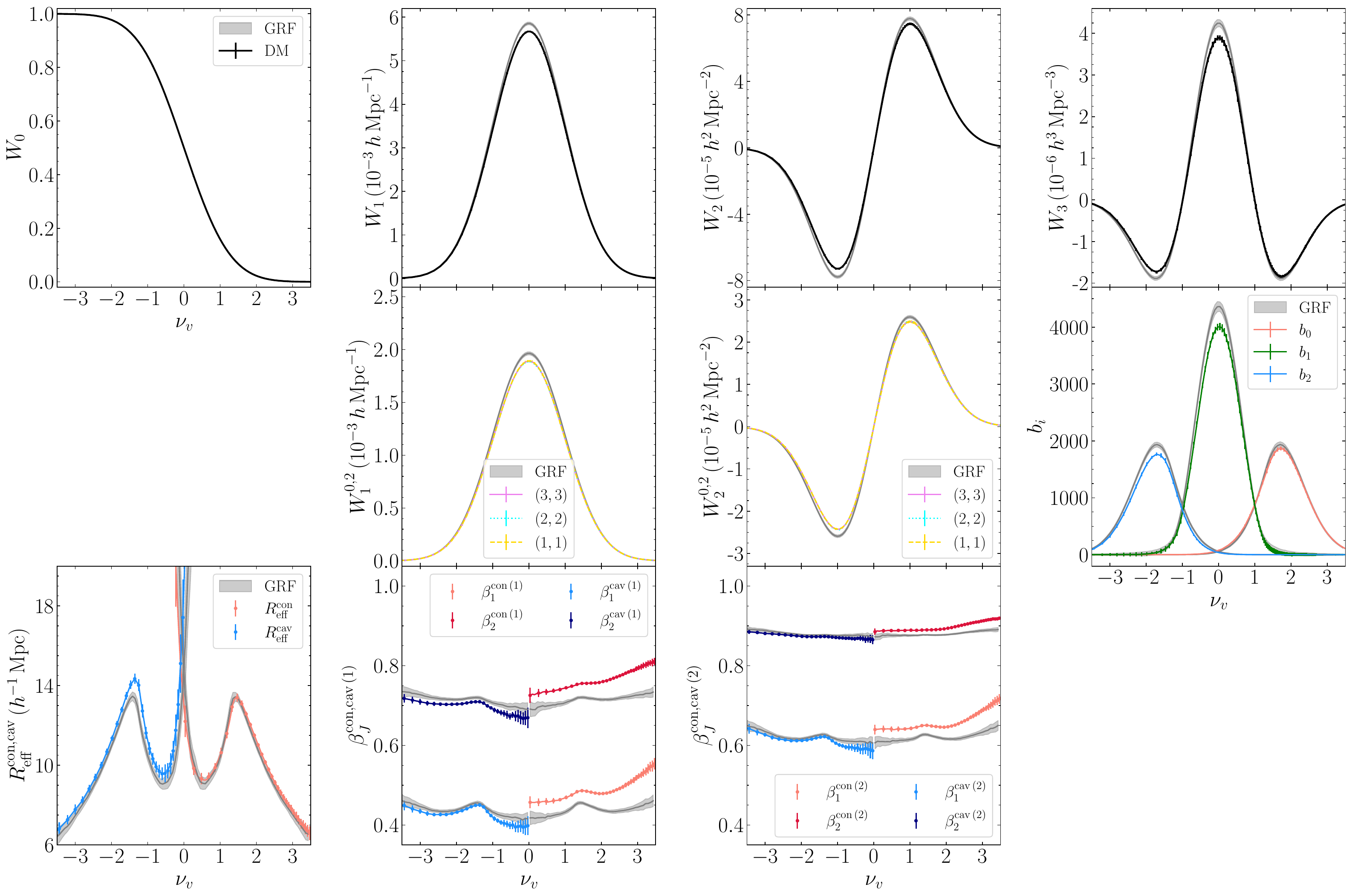}
    \caption{Identical to Figure~\ref{fig:2_global}, except that the coloured dark matter curves are presented as functions of volume threshold $\nu_{v}$ as opposed to conventional iso-field values $\nu$. The grey Gaussian reference curves are unaffected by the $\nu \to \nu_{v}$ transformation. 
     } 
    \label{fig:2_global_nuv}
\end{figure}

Figure~\ref{fig:2_global} presents a direct comparison between Gaussian and non-Gaussian fields at equal iso-field thresholds $\nu$. Although this comparison is well defined, it is not the most suitable choice for understanding the effect of gravity as manifested in the morphological statistics. 
The reason is that since gravitational evolution strongly skews the PDF of the density field, $P(\delta)$, a fixed threshold $\nu$ corresponds to different quantiles of the Gaussian and non-Gaussian PDFs. In other words, for the same value of $\nu$, the two excursion sets of the Gaussian and gravitationally evolved density field sample different regions of the underlying  PDFs.

One way to overcome the issue of quantile mismatch  and to meaningfully compare the morphology of Gaussian and non-Gaussian fields is to express morphological statistics as functions of  the  
volume threshold $\nu_{v}$ defined by \citep{1987ApJ...319....1G, 1987ApJ...321....2W,1988ApJ...328...50M}: 
\begin{equation} f_{v} = {1 \over \sqrt{2\pi}} \int^{\infty}_{\nu_{v}} e^{-t^{2}/2} dt , \end{equation} 
where $f_{v}$ is the fractional volume of the field above $\nu_{v}$~\citep{1987ApJ...319....1G}. 
For each given field, we fix a value of $f_v$ and use it to determine $\nu_v$ (the mapping is one-to-one). We then consider the morphological statistics as functions of $\nu_v$. A  demonstration of this mapping is shown in Figure~\ref{fig:nu_nuv} in Appendix~\ref{app:a}. For a Gaussian field, $f_{v}$ is exactly $W_0$, and $\nu_v$ is identical to $\nu$. However, for a gravitationally evolved non-Gaussian field, $\nu_v\ne \nu$ (see Figure~\ref{fig:nu_nuv} for their exact relation).  Therefore, comparing excursion sets of the Gaussian and non-Gaussian fields by fixing $\nu_v$  corresponds to comparing the two fields at the same volume fraction, or equivalently at the same  quantile of the field distribution. Using $\nu_{v}$ exactly transforms the one-point PDF of the field $\delta$ to its Gaussian form. 

The impact of using the volume threshold can be understood as follows. The volume fraction of the excursion set gives the cumulative one-point probability of the field, assuming ergodicity. Hence only $P(\delta)$ is required to determine its statistical properties. In contrast, other morphological statistics (higher order MFs, Betti numbers and associated local statistics) are determined not just by $P(\delta)$, but by the joint PDF of $\delta$ and its higher derivatives. 
Note that the derivative of the field along the $i^{\rm th}$ coordinate can be written as 

\begin{equation} \partial_{i} \delta = \lim_{\epsilon \to 0} {1 \over \epsilon}
\left[\delta(x_{i} + \epsilon) - \delta(x_{i})\right] , \end{equation} 

so that all correlations between the field at $x_{i}$ and $x_{i}+\epsilon$, whose ensemble average is not zero in the limit $\epsilon \to 0$, are encoded in the derivatives of the field. Using the volume threshold then amounts to removing  the purely local, one-point contribution to the non-Gaussianity, while retaining all the `non-local' information contained in the joint PDF of the field and its derivatives. Since we wish to isolate the impact of gravitational evolution on the spatial correlations of the field, we  present our measurements of morphological statistics of the dark matter fields, as functions of $\nu_{v}$, in Figure~\ref{fig:2_global_nuv}. The colour scheme matches that of Figure~\ref{fig:2_global}. 

In the first row of Figure~\ref{fig:2_global_nuv}, we compare the Minkowski functional curves computed from the gravitationally evolved non-linear matter density field and those from an initial GRF (shown as black and grey curves, respectively). For the $W_{1}$, $W_{2}$, and $W_{3}$ curves, we observe a noticeable suppression of the amplitudes in the dark matter measurements. This is the well-known {\it gravitational smoothing} effect \citep{1988ApJ...328...50M, Kim_2014},
which can be described using the four-point function on quasi-linear scales \cite{Matsubara:2020knr}. There is also some residual skewness in the curves, which is most apparent in $W_{3}$, with negative thresholds more significantly impacted by the gravitational evolution. 

Although non-linear gravitational evolution introduces significant non-Gaussianity, it preserves statistical isotropy. This property is verified by the globally averaged Minkowski Tensors $W_{1}^{0,2}$ and $W_{2}^{0,2}$, shown in the middle panels of Figure~\ref{fig:2_global_nuv}. The diagonal components of these rank-2 tensors, $W_{1}^{0,2}|_{\mu}^{\mu}$ and $W_{2}^{0,2}|_{\mu}^{\mu}$ (where $\mu=1, 2, 3$), overlap exactly for the non-linear matter field, while the off-diagonal elements (not shown) remain consistent with zero. Locally, we expect gravitational collapse to occur anisotropically, with matter collapsing onto wall saddle points from underdense regions, finally fed into peaks via filaments. However, there is no preferred global orientation in these locally anisotropic processes, and the MTs are constructed in a coordinate system defined over the entire domain. Provided the spatial averaging is performed over volumes much larger than the characteristic scale of individual structures, gravitational collapse preserves statistical isotropy as quantified by the Minkowski tensors.

The remaining four panels of the Figure~\ref{fig:2_global_nuv}: the Betti curves (second row, right panel) and the individual structure morphology (bottom-row panels) illustrate how gravitational collapse affects the connected components and cavities. The blue curves inform us that in the $z=0$ dark matter field there are fewer but larger cavities compared to the Gaussian reference. The change in number and size of cavities is likely due to the merging of minima critical points through the consumption of wall saddles. In contrast, $b_{0}$ (red curve) informs us that there are slightly fewer connected components, but their size $R_{\rm eff}^{{\rm con}}$ remains indistinguishable from that in the Gaussian field. The decrease in $b_{2}$ is significantly larger than $b_{0}$, which indicates that minima-wall saddle merging is a more efficient process than peak-filament merging on these large scales. 

The shape of the cavities, characterized by $\beta_{J}^{{\rm cav} \, (1,2)}$ (blue curves, bottom-row middle and right panels) is marginally less spherical than the Gaussian counterparts, but otherwise practically unchanged. The shape of the connected components (red curves) evolve more significantly, becoming more spherical for all thresholds probed. The degree of sphericalization increases with increasing $\nu_{v}$. We also note that the peaks of the curves in the bottom panels remain located at $\nu_{v} \simeq 1.4$, unaffected by the gravitational evolution after the $\nu \to \nu_{v}$ threshold re-mapping. 

The overall effect of gravitational collapse on these quasi-linear scales is to merge a fraction of minima via the consumption of wall saddle points, thereby increasing their average size, while simultaneously making the peaks more spherical. The Minkowski Functionals and tensors exhibit reduced amplitudes and a slight skewness relative to the Gaussian case. 

\section{Redshift Space Distortion} 
\label{sec:rsd}

Next, we consider the effect of redshift-space distortion (RSD) on excursion-set morphology. This phenomenon corresponds to the anisotropic deformation of observed structures caused by contamination of cosmological redshifts by line-of-sight peculiar velocities. The physical anisotropic effects of RSD can be understood as two distinct phenomena. The first is the `Kaiser effect' in which coherent peculiar velocities cause large-scale structures to appear compressed along the line of sight~\cite{1987MNRAS.227....1K,1998ASSL..231..185H}. This gravitational infall is observed on all scales. The second is the `Finger-of-God effect'~\cite{1972MNRAS.156P...1J} on smaller scales due to large random velocities of virialised objects which stretch structures along the line of sight, producing elongations pointing toward the observer. 

Under the plane-parallel approximation, redshift-space density field, $\delta({\bf s})$, can be generated from the real-space density field by perturbing particle positions $\mathbf{x}$ along the line-of-sight direction. If the line of sight is taken to be the $x_{3}$-axis (the ${\bf e}_3$ direction), the displacement is determined by the particle peculiar velocity $\mathbf{v}$ such that:

\begin{equation}
{\bf s}= {\bf x}+{\bf e}_{3}( {\bf v} . {\bf e}_{3})\frac{1+z}{H(z)}.
\end{equation}
Because RSD explicitly breaks statistical isotropy, we expect shapes and sizes of excursion sets to differ between directions parallel and perpendicular to the line of sight. In this section, we investigate these anisotropic effects. 

\subsection{Linear RSD}
Before we study the full, non-linear effect of redshift-space distortion on the evolved $z=0$ dark matter particle distribution, we first consider the linear regime. We generate linear redshift-space distorted fields by taking the isotropic Gaussian random fields $\delta^{(r)}({\bf x})$ constructed in Section~\ref{sec:grf} and applying the following transformation in Fourier space~\cite{1987MNRAS.227....1K} : 

\begin{equation} \delta^{(s)}({\bf k}) = (1 + f \mu^{2}) \delta^{(r)}({\bf k})
\end{equation} 
where $\mu = {\bf e}_{3} . {\bf k} /|{\bf k}|$ and $f = \Omega_{m}^{\gamma}$ with $\gamma \simeq 0.55$. The superscripts $(r)$ and $(s)$ denote real and redshift space, respectively.  After performing the inverse Fourier transform, the resulting field contains the anisotropic Kaiser signal but does not include the non-linear Finger-of-God contribution or higher-order anisotropic mode couplings. Because the mapping between real- and redshift-space fields is linear, and the RSD operator is deterministic, the field $\delta^{(s)}({\bf x})$ remains Gaussian. 

\begin{figure}[h!]
    \centering
    \fbox{\large Gaussian Fields, Redshift Space}\\
    \vspace{.2cm}
    \includegraphics[width=0.98\textwidth]{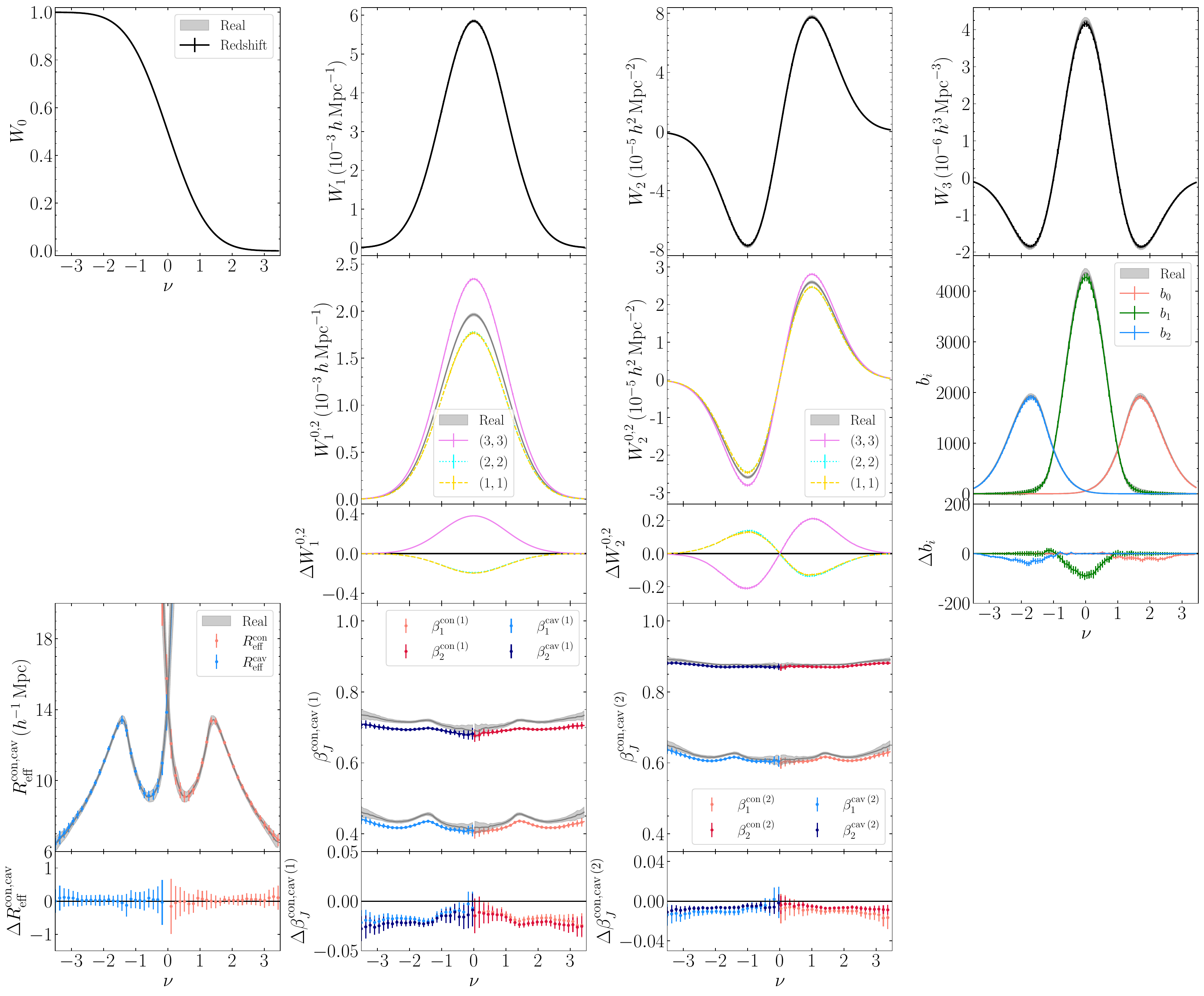}
    \caption{Morphological statistics measured for Gaussian fields in linear redshift space. Coloured curves show the redshift-space results, while the grey curves correspond to the isotropic Gaussian case from Figure~\ref{fig:1_global} and are included for comparison. The panel layout and colour scheme follow Figure~\ref{fig:1_global}, except that the second and third rows include additional sub-panels showing the difference between the redshift- and real-space statistics.}  
    \label{fig:2_global_grf}
\end{figure}

We generate $N_{\rm real} = 50$ anisotropic Gaussian random fields using the Quijote fiducial cosmology and a smoothing scale $R_{G} = 10 \, h^{-1} \, {\rm Mpc}$, and extract the global and local summary statistics, which are presented in Figure~\ref{fig:2_global_grf}. The color scheme matches that of Figure~\ref{fig:2_global_nuv}.  In each panel, the grey filled lines/bands represent the mean and uncertainty of the same statistics extracted from the corresponding real-space Gaussian fields. 

In the top panels of Figure~\ref{fig:2_global_grf}, we see that Minkowski Functionals of the redshift-space fields (black curves) are practically indistinguishable from the corresponding real-space grey curves, with only a very small ($\sim 1\%$) decrease in their amplitudes in redshift space. The largest amplitude difference occurs for $W_{3}$, which has been calculated analytically in \cite{1996ApJ...457...13M}. In the right panel of the second row, we present the Betti numbers, whose amplitudes are all marginally lower in redshift space, commensurate with $W_3$, in redshift space. In the sub-panel directly beneath, we present the difference between the redshift- and real-space Betti numbers together with the error on the mean of the measurement. The suppression is approximately $2\%$. 

In the second row of Figure~\ref{fig:2_global_grf}, the middle panels display the Minkowski tensor components. We plot the components of $W_{1}^{0,2}$ and $W_{2}^{0,2}$ , as well as the differences $\Delta W^{0,2}_{i} = W^{0,2}_{i}({\rm redshift}) - W^{0,2}_{i}({\rm real})$ for each principal axis $(1,1), (2,2), (3,3)$. The pink dashed curve corresponds to the tensor component oriented parallel to the line-of-sight, while the yellow and cyan curves are the two degenerate components perpendicular to the line-of-sight. The parallel and perpendicular components deviate significantly, demonstrating that the MTs capture the strong anisotropy imprinted by RSD on the matter density field. For the Gaussian fields considered in this subsection, this difference corresponds to a constant amplitude offset. In contrast to the scalar MFs, we find that MT components clearly capture the global anisotropy introduced by linear redshift-space distortions.

In the third row (left panel), we find that the effective radii $R_{\rm eff}^{{\rm con},{\rm cav}}$ of connected components and cavities are practically unaffected by the linear redshift-space distortion operator. However, their shapes (middle and right panels) become systematically less spherical, i.e. more anisotropic, at all thresholds. In the lower sub-panels we present the difference between redshift- and real-space statistics. While the radii exhibit no measurable change, we observe decrease of $\sim 0.03$ and $\sim 0.01$  in $\beta_{J}^{{\rm con},{\rm cav} \, (1)}$ and $\beta_{J}^{{\rm con},{\rm cav} \, (2)}$, respectively. These values indicate that $\beta_{J}^{{\rm con},{\rm cav} , (1)}$ responds more strongly to linear RSD than $\beta_{J}^{{\rm con},{\rm cav} , (2)}$, implying that the presence of the mean curvature in the integrand of $W_2^{0,2}$ reduces its sensitivity to anisotropy compared to $W_1^{0,2}$. 

In conclusion, while individual objects in real space are randomly oriented with respect to the line-of-sight ${\bf e}_{3}$,  in redshift space their major axes have a tendency to be preferentially aligned with ${\bf e}_{3}$. The distortion effect is small but coherent, and is responsible for the systematic decrease in $\beta_{J}^{{\rm con},{\rm cav} \, (1)}$ and $\beta_{J}^{{\rm con},{\rm cav} \, (2)}$.

\subsection{Non-Linear RSD}
\begin{figure}
    \centering
    \fbox{\large Gravitationally Evolved Dark Matter, Redshift Space}\\
    \vspace{.2cm}
    \includegraphics[width=0.98\textwidth]{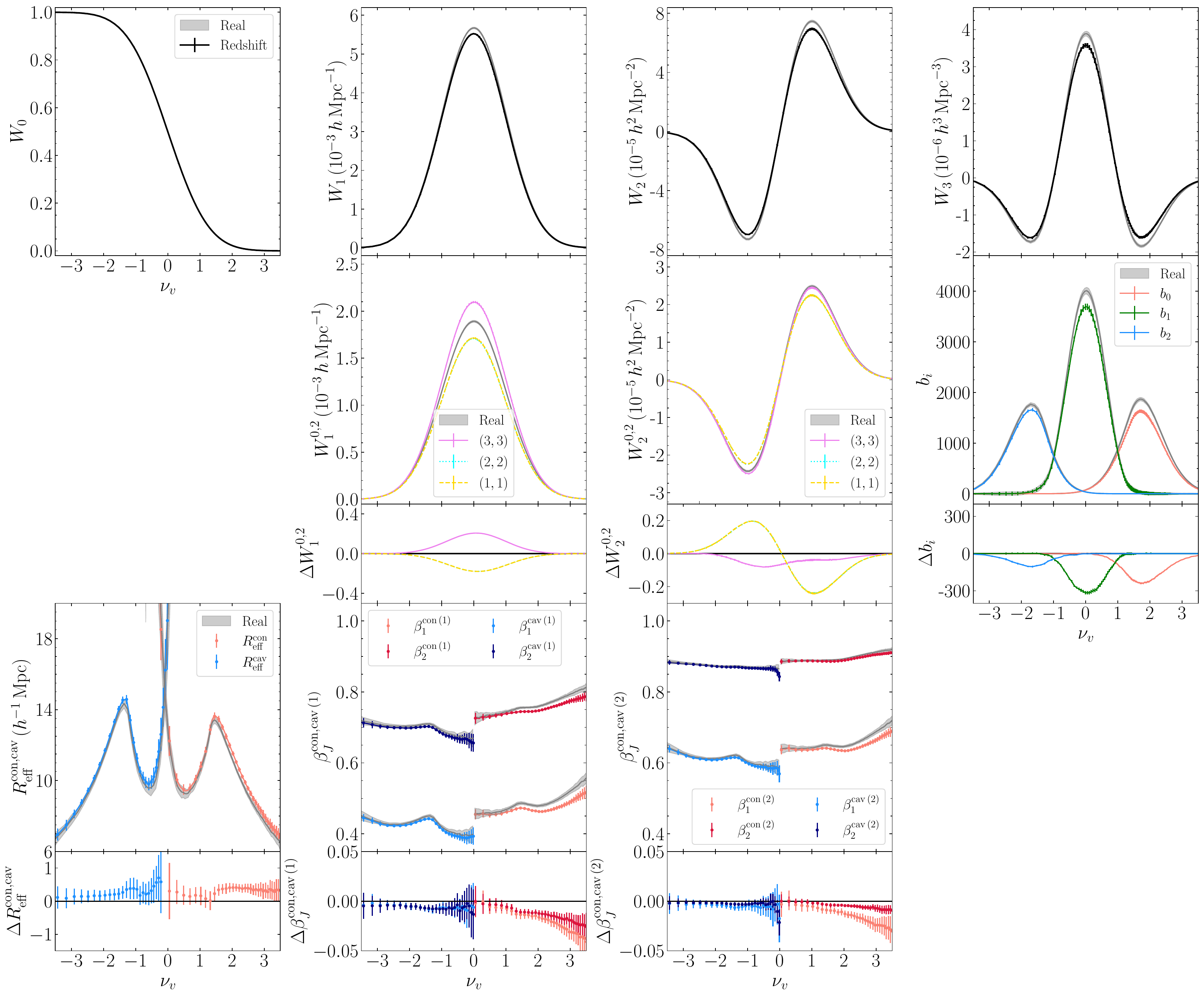}
     \caption{Morphological statistics measured for dark matter fields in non-linear redshift space. Coloured curves show the redshift-space results, while the grey curves correspond to real space dark matter from Figure~\ref{fig:2_global_nuv} and are included for comparison. The panel layout and colour scheme follow Figure~\ref{fig:2_global_grf}, including the sub-panels showing the difference between redshift and real space statistics.}
    \label{fig:3_global}
\end{figure}

When we directly correct dark matter particle positions using their peculiar velocity along the line of sight, and use the new positions to construct a density field in redshift space, we are accounting for the full non-linear redshift-space distortion. This includes the higher order non-Gaussianty from the Kaiser effect as well as the non-linear Finger-of-God (FoG) stochastic velocity dispersion. Figure~\ref{fig:3_global} compares the morphological statistics of the non-linear dark matter field in real space (grey curves) and redshift space (coloured curves). The colour scheme is the same as in previous figures.

In the top row, we find that compared to linear RSD summarized in the previous subsection, there is a larger decrease in amplitude of the MFs in redshift space compared to real space. This stronger suppression is due to the FoG dispersion along the line of sight, which washes out small-scale structures by acting as an additional source of anisotropic smoothing. Furthermore, it generates negative skewness, most clearly observed as asymmetry of $W_{3}$ (right panel). FoG affects overdense structures, leaving a larger imprint on  $\nu_{v} > 0$ thresholds. This behaviour is also reflected in the Betti curves (second row, third panel), where the amplitude of  $b_{0}$ (red curve) decreases more strongly compared to $b_{2}$. The processes of gravitational evolution and non-linear RSD act contrarily on the MFs in the following sense - the former primarily decreases the number of minima via merging, while the latter merges peaks along the line of sight via FoG smoothing. This makes the redshift-space statistics less skewed than their real-space counterparts. However, the two effects do not cancel one another, because both also reduce the overall amplitude of the Minkowski Functionals and Betti numbers. 

The Minkowski Tensors (second row, first two panels) still contain the Kaiser signal, but the line-of-sight components are modified due to FoG. The yellow and cyan curves, representing the components perpendicular to the line of sight, present almost the same signal as their linear RSD counterparts in Figure~\ref{fig:2_global_grf}, specifically a constant amplitude decrease in redshift space. Contrarily, the line-of-sight-parallel curves (pink) are strongly modified in both amplitude and shape by non-linear effects. In particular, the amplitude of the $(3,3)$ component of $W_{1}^{0,2}$ is reduced compared to the linear RSD signal, and for $W_{2}^{0,2}$ the shape of the $(3,3)$ curve is also modified. This is most clearly seen in the difference sub-panels, which again show $\Delta W^{0,2}_{i} = W^{0,2}_{i}({\rm redshift}) - W^{0,2}_{i}({\rm real})$, but now for the non-Gaussian fields. 

We also consider how non-linear RSD influences the size and shape of individual structures. In the third row, left panel of  Figure~\ref{fig:3_global}, we present $R^{{\rm con}}_{\rm eff}$ for connected components (red curves) and $R^{{\rm cav}}_{\rm eff}$ for cavities (blue curves). In the sub-panel below, we show the differences $\Delta R_{\rm eff}^{{\rm con}, {\rm cav}} = R_{\rm eff}^{{\rm con}, {\rm cav}}({\rm redshift}) - R_{\rm eff}^{{\rm con}, {\rm cav}}({\rm real})$. The overall effect is a small increase in effective radius, that is only significant in $R_{\rm eff}^{{\rm con}}$ at relatively high thresholds $\nu_{v} > 1.5$. 

The lower middle and right panels of Figure~\ref{fig:3_global} show the shape anisotropy parameters $\beta^{{\rm con}, {\rm cav} (1)}_{J}$ and $\beta^{{\rm con}, {\rm cav} (2)}_{J}$. For the linear redshift-space distorted fields discussed in the previous section, both connected components and cavities exhibited lower values of $\beta^{{\rm con}, {\rm cav} (1)}_{J}$ and $\beta^{{\rm con}, {\rm cav} (2)}_{J}$ compared to their real-space counterparts. In contrast,  distortions arising from non-linear RSD shift the average shape of cavities (blue curves) closer to their isotropic values, whilst increasing the ellipticity of high-threshold connected components (red curves). This trend can be understood physically: virialised motions within overdense regions generate strong FoG elongations along the line of sight, whereas underdense regions are less affected by such stochastic velocities. Once again, the effects of gravitational evolution and non-linear RSD act contrarily on the statistics : gravity slightly sphericalizes the connected components, while RSD elongates them. The two competing effects are not degenerate in the $\beta_{J}^{{\rm con}, {\rm cav} \, (1)}$ and $\beta_{J}^{{\rm con}, {\rm cav} \, (2)}$ curves, with redshift-space distortions inducing a smaller overall impact.

In summary, among the various phenomena studied in this work, RSD induces the most pronounced morphological changes in the Minkowski Tensors $W_{1}^{0,2}$ and $W_{2}^{0,2}$, and both linear and non-linear processes can be studied through their behaviour. Both gravitational evolution and RSD act to decrease the amplitude of the scalar MFs, while the two processes skew the MF curves in opposite directions. The shape of connected components, as quantified by $\beta_{J}^{{\rm con} \, (1)}$, are most significantly altered by non-linear RSD effects, producing a threshold-dependent elongation of the excursion sets that is largest for isolated high-density peaks. 

\section{Neutrino Mass}
\label{sec:neutrinos}

Next, we consider the impact of massive neutrinos on excursion-set morphology by comparing simulations with and without a massive neutrino component (specifically $M_{\nu}=0.4\, {\rm eV}$ vs. $M_{\nu}=0$). We direct the reader to ~\citep{Villaescusa-Navarro:2019bje} for details on the implementation of massive neutrinos within the simulations, and \cite{Liu:2022vtr,Liu:2023qrj,Liu:2024uxa} for a detailed study of the impact of neutrino mass on the Minkowski Functionals. When massive neutrinos are present as an additional energy component, there is an ambiguity in the definition of the dark matter field, specifically whether the massive neutrino particles are included in the definition of $\delta({\bf x})$. In this work, density fields are constructed from  dark matter particles only, excluding the neutrinos, since galaxy catalogues are expected to trace the clustered dark matter component. 
The mass value $M_{\nu} = 0.4 \, {\rm eV}$ used in this work lies close to the current model-independent upper bound on the effective electron-flavour neutrino mass from the KATRIN experiment, 
$m_{\beta}<0.45$ eV at 90\% confidence level \cite{katrin:2025}. When expressed in terms of the total neutrino mass, current cosmological analyses yield upper limits in the range $\Sigma m_\nu \lesssim 0.07$-$0.12\,\mathrm{eV}$ (95\% C.L.), with the precise bound depending on the combination of datasets employed. The strongest constraints arise from analyses that include the latest baryon acoustic oscillation (BAO) data from the Dark Energy Spectroscopic Instrument (DESI), which substantially enhance sensitivity to neutrino mass~\cite{Aghanim:2018eyx,2025JCAP...02..021A}. 

\begin{figure}
    \centering
    \fbox{\large Gravitationally Evolved Dark Matter with Massive Neutrinos}\\
    \vspace{.2cm}
    \includegraphics[width=0.98\textwidth]{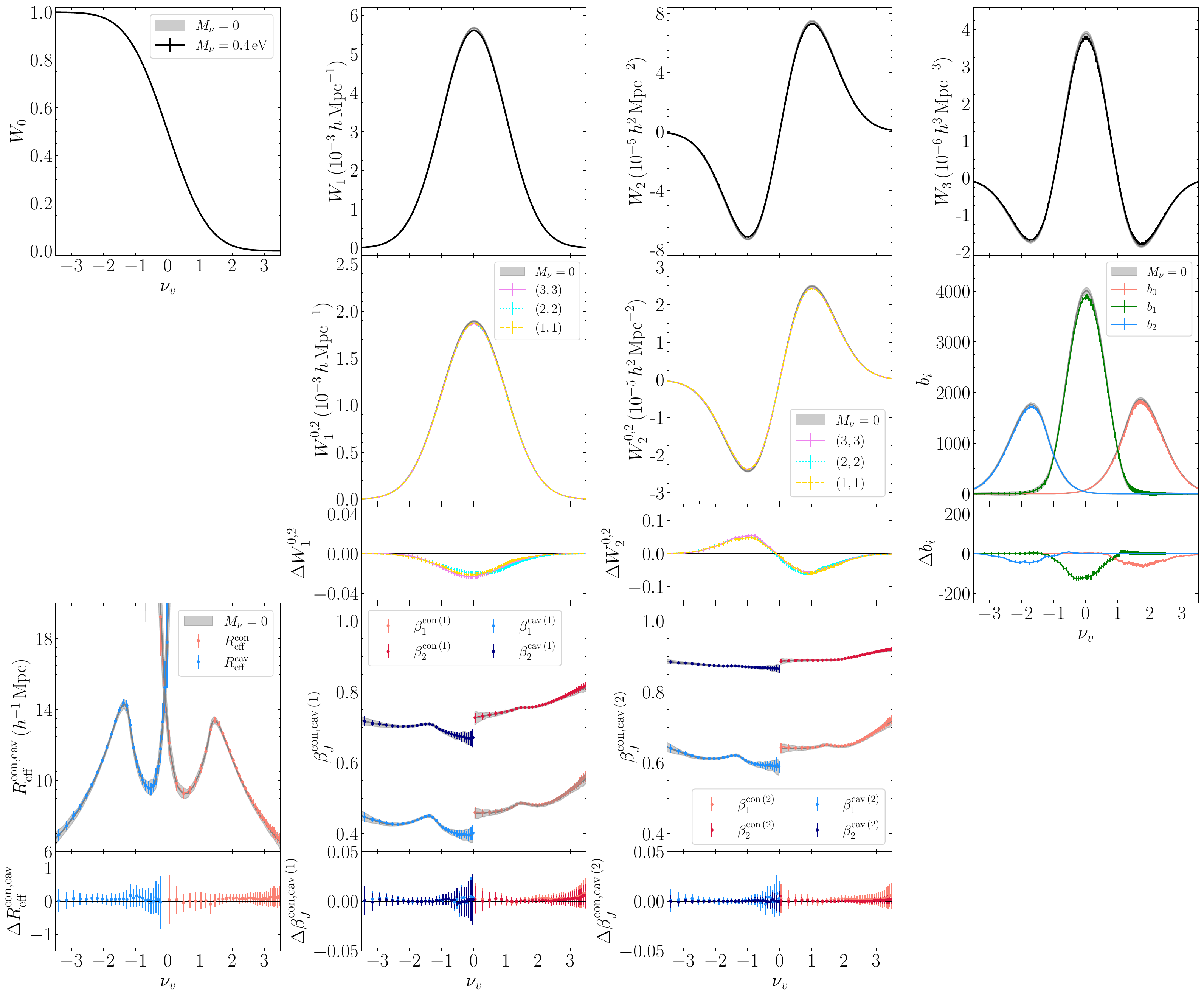}
    \caption{Morphological statistics measured for dark matter fields in real space including a massive neutrino component. Coloured curves show the massive-neutrino case, while the grey curves correspond to the massless-neutrino results from Figure~\ref{fig:2_global_nuv} and are included for comparison. The panel layout and colour scheme follow Figure~\ref{fig:3_global}, including subpanels that show the difference between the massive- and massless-neutrino cases.}
    \label{fig:4_global}
\end{figure}

In Figure~\ref{fig:4_global}, we present plots of our suite of summary statistics extracted from $N_{\rm real} = 50$, $M_{\nu} = 0.4 \, {\rm eV}$ snapshot boxes at $z=0$ (coloured curves) and the corresponding dark matter snapshot boxes with $M_{\nu} = 0 $ (grey curves). The comparison is made in real space in order to disentangle the effects of RSD and massive neutrinos. 

We first note that in all panels the impact of massive neutrinos is small and barely discernable by eye. In fact, the only effect is a slight decrease of the amplitudes of the MFs $W_{1,2,3}$, MTs $W_{1}^{0,2}$, $W_{2}^{0,2}$ and the Betti numbers. The amplitude drop is approximately $\sim 1-3\%$, and is largest for $W_{3}$ and the Betti numbers. This arises from the well-known damping of small-scale structures by neutrino free-streaming: because massive neutrinos fail to cluster efficiently below their free-streaming scale, they smoothen the high-contrast in the density field, reducing both the curvature and the total surface area of excursion-set boundaries. Although the damping effect is most pronounced on small scales $\lesssim 1 \, h^{-1} \, {\rm Mpc}$, there is a small reduction in clustering at $\sim 10 \, h^{-1} \, {\rm Mpc}$, as we see here. The effect is isotropic, as expected, and evidenced by the MTs remaining proportional to the identity matrix.   

Further, massive neutrinos do not significantly modify the shape or size of individual components. The effective radius curves $R_{\rm eff}^{{\rm con}, {\rm cav}}$ and the shape parameters $\beta_J^{{\rm con}, {\rm cav} \, (1), (2)}$ for  $M_{\nu}=0.4$ eV and $M_{\nu}=0$ remain within the error on the mean of our measurements (Figure~\ref{fig:4_global}, bottom panels). The shape parameters, $\beta_J^{{\rm con}, {\rm cav} \, (1)}$ and $\beta_J^{{\rm con}, {\rm cav} \, (2)}$, are nearly identical in the massive- and massless-neutrino case. This is expected as neutrino free-streaming damps fluctuations isotropically, reducing small-scale contrast without introducing any directional distortion, and thereby leaving the shape parameters essentially unchanged ($\Delta \beta_J^{{\rm con}, {\rm cav} \, (1), (2)} \sim 0$).

The imprint of neutrinos on large-scale structures is subtle, and is predominantly observed on small scales. When extracting the Minkowski functionals from galaxy surveys, it is common to smooth the point distribution with a constant comoving scale to construct a continuous density field\footnote{However, topology can be extracted directly from point distributions without generating a continuous field \cite{Pranav:2016gwr,10.1111/j.1365-2966.2011.18394.x,Aggarwal:2024yqh}}. To minimize shot noise, a smoothing scale of $R_{\rm G} \sim \bar{r} = \bar{n}^{-1/3}$ is required, where $\bar{n}$ is the number density of the points. This requirement restricts us to large scales, as current generations of galaxy catalogues typically have $\bar{r} \simeq 10 \, h^{-1} \, {\rm Mpc}$. To explore the effect of neutrinos, we will likely have to introduce alternative approaches, such as adaptive smoothing scales that preserve small-scale information. 

\section{Parameter Sensitivity}
\label{sec:cosparams}

Finally, for the dark matter fields in real space, we explore the sensitivity of the global and local statistics to cosmological parameters. The Minkowski Functionals are finding increasing application in cosmology~\cite{Hikage:2006fe,2016MNRAS.461.1363N,2017ApJ...836...45A,Chingangbam:2017PhLB,Marques:2018ctl,Appleby:2020pem,Rahman:2021,Appleby:2021xoz,Liu:2022vtr,Liu:2023qrj,2024JCAP...01..036R}, and here we consider the response of a broader family of summary statistics to the cosmological parameters $\Omega_{m}$, $n_{s}$ and $\sigma_{8}$. We restrict our analysis to this subset, assuming that tighter constraints on $\Omega_{b}$ and $h$ will arise from complementary cosmological data. 

To proceed, we use $N=50$ realisations of the Quijote simulations for a set of different cosmological models, the parameters of which are presented in Table~\ref{tab:q_columns}. To estimate the parameter sensitivity, we first define the derivative of each statistic $Y_k$, at each threshold and for each realization, using the standard finite-difference scheme: 
\begin{equation}
   \label{eq:sens} {\partial Y_{k} \over \partial \theta} =
  {Y_{k}(\theta^{+}) - Y_{k}(\theta^{-}) \over \theta^{+} - \theta^{-}},
\end{equation}
\noindent where $\theta^{\pm}$ are the values of the parameters listed in Table~\ref{tab:q_columns}. In this section, we revert to the iso-field threshold $\nu$, rather than $\nu_{v}$. We worked with $\nu_{v}$ threshold in the previous sections to qualitatively study the response of morphology and topology to various phenomena, effectively removing the skewness of the purely local one-point function. In this section, we  however aim to quantify the information content of the summary statistic curves in the Fisher information sense. Since the one-point skewness itself is sensitive to cosmological parameters, and we seek to maximize the parameter sensitivity, we return to the $\nu$ parameterization. 

\begin{figure}
    \centering
    \includegraphics[width=0.98\textwidth]{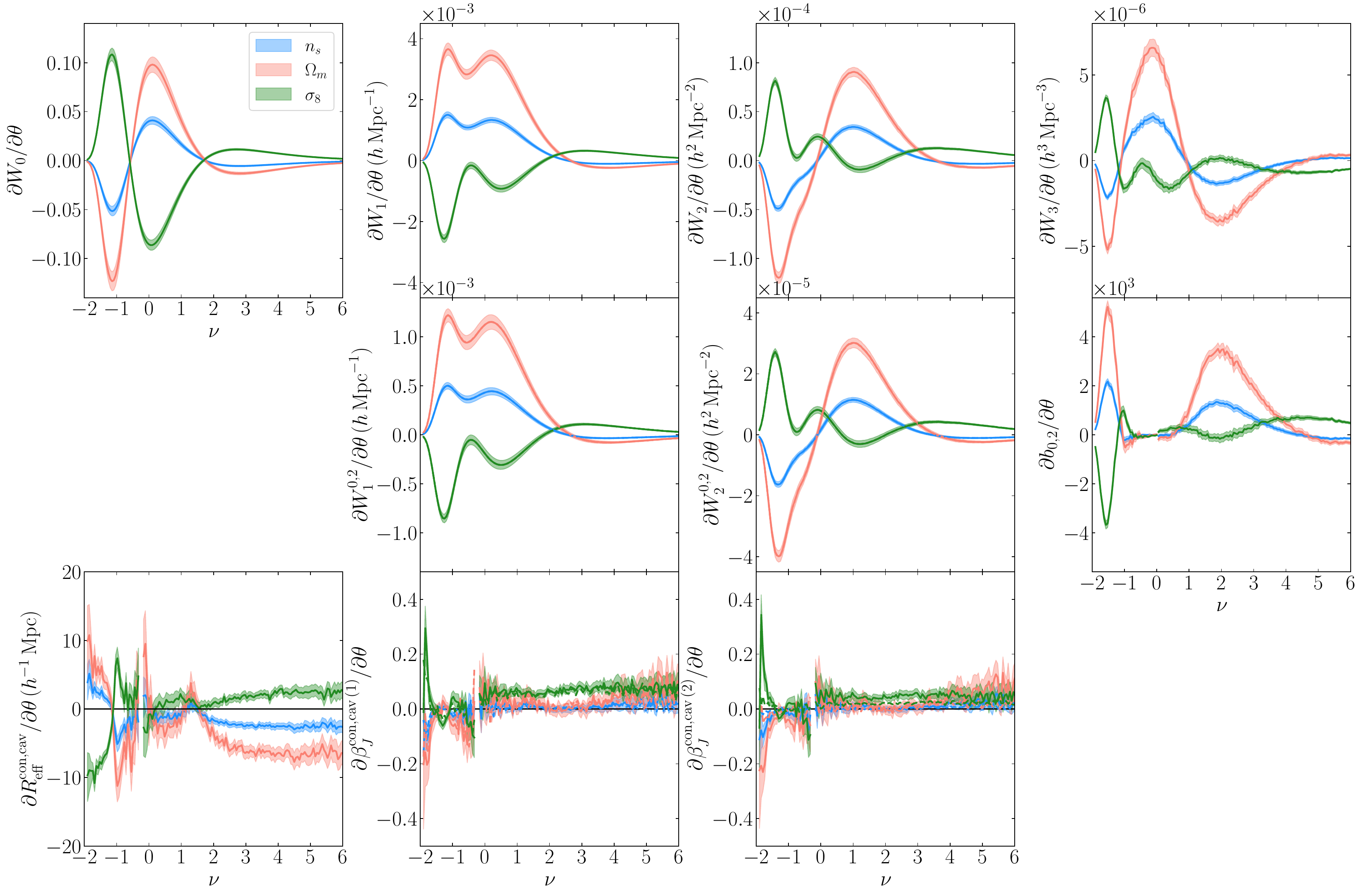}
    \caption{Sensitivity of the statistics to cosmological parameters $n_{s}$, $\Omega_{m}$ and $\sigma_{8}$ (blue/red/green bands). The lines/filled regions represent the mean/error on the mean of $N=50$ realisations of each parameter set. }
    \label{fig:parameters_1}
\end{figure}

The results for the global and local statistics are shown in Figure~\ref{fig:parameters_1}. The blue/red/green lines with shaded regions represent the mean and the error on the mean of the sensitivities to $n_{s}$, $\Omega_{m}$ and $\sigma_{8}$, respectively. In the middle panels of Figure~\ref{fig:parameters_1}, which show $\partial W_{1}^{0,2}/\partial \theta$ and $\partial W_{2}^{0,2}/\partial \theta$, there are three distinct line styles (solid, dashed, dotted) for each colour. These correspond to the three diagonal components of the Minkowski tensors, but are not visually distinguishable as they overlap within the error bars, as expected for statistically isotropic fields. 

In the top panels of Figure~\ref{fig:parameters_1}, the blue and red curves differ primarily by an overall amplitude, indicating a strong degeneracy between $n_{s}$ and $\Omega_{m}$. This is a generic property of the Minkowski Functionals : their shapes are largely determined by a small number of one-point cumulants on quasi-linear scales $R_{G} \simeq 10 \, h^{-1} \, {\rm Mpc}$. These cumulants depend on the shape of the matter power spectrum, and over a limited range of scales, variations of $n_{s}$ and $\Omega_{m}$ can mimic one another, creating the observed degeneracy. There are two potential ways to break this degeneracy. The first is to combine measurements on relatively small ($R_{G} \sim 10 \, h^{-1} \, {\rm Mpc}$) and large ($R_{G} \sim 30\, h^{-1} \, {\rm Mpc}$) smoothing scales, although the latter would suffer more from cosmic variance. The second is to adaptively smooth the field using a multi-scale kernel, which preserves the small scale information and probe the field in a regime in which perturbation theory breaks down. 

In the Gaussian limit, the Minkowski functionals are independent of $\sigma_8$, which merely scales the amplitude of density fluctuations. By contrast, for gravitationally evolved, non-Gaussian fields, variations in $\sigma_8$ do more than rescale the amplitude; they also encode the degree of non-Gaussianity in the field. This difference explains the qualitatively distinct behaviour of the green curves relative to the blue and red ones. The effect is most clearly seen in the derivatives $\partial W_{1,2,3}/\partial \theta$ with respect to $n_s$ and $\Omega_m$, which include a term proportional to $W_{1,2,3}$. This proportionality indicates that these derivatives retain a substantial fraction of Gaussian information. In contrast, 
$W_0$ carries no cosmological information in its Gaussian expectation value, and therefore the curves for $\partial W_{0} /\partial \theta$ shown in the top-left panel arise entirely from non-Gaussian contributions. 

In the second row of Figure~\ref{fig:parameters_1}, the first two panels present the parameter sensitivity of the Minkowski Tensors. For each colour, the three curves (solid, dotted, dashed) represent the $(1,1)$, $(2,2)$, $(3,3)$ diagonal components, but they overlap entirely. The shapes of these curves match the corresponding MF curves above, indicating that no additional information is gained from the tensors when the field is isotropic. The Betti curves, shown in the right panel of the second row, do exhibit some additional discriminatory power. To avoid clutter, we only present $\partial b_{2}/\partial \theta$ and $\partial b_{0}/\partial \theta$ for negative/positive thresholds respectively, and omit $\partial b_{1}/\partial \theta$. The coloured curves show some variation in shape; for example, $\partial b_{2}/\partial \sigma_{8}$ (cf. green curve at negative thresholds) shows a secondary bump at $\nu \sim -1$ which is not discernable in the other cases. 

\begin{table}[h!]
\centering
\begin{tabular}{| c c c c |}
\hline
Parameter   & $\Omega_{m}$        & $\sigma_{8}$    & $n_{s}$   \\ \hline
Fid & 0.3175   & 0.834  & 0.9624 \\
$\Omega_{m}^{+}$ & \bf{0.3275}  & 0.834  & 0.9624 \\ 
$\Omega_{m}^{-}$ & \bf{0.3075}  & 0.834  & 0.9624 \\ 
$\sigma_{8}^{+}$ & 0.3175  & \bf{0.849}  & 0.9624 \\ 
$\sigma_{8}^{-}$ & 0.3175  & \bf{0.819}  & 0.9624 \\ 
$n_{s}^{+}$ & 0.3175  & 0.834  & \bf{0.9824} \\ 
$n_{s}^{-}$ & 0.3175  & 0.834  & \bf{0.9424} \\ 
\hline 
\end{tabular}
\caption{Cosmological Models used in this work. Other cosmological parameters are fixed as $\Omega_{b}=0.049$, $h = 0.6711$. Bold values indicate the parameter being varied from the fiducial.}
\label{tab:q_columns}
\end{table}

In the bottom panels of Figure~\ref{fig:parameters_1}, we present the parameter sensitivity of the local statistics $R_{\rm eff}^{{\rm con}, {\rm cav}}$, $\beta_{J}^{{\rm con}, {\rm cav} \, (1)}$, $\beta_{J}^{{\rm con}, {\rm cav} \, (2)}$. For clarity, we show these quantities for $\nu > 0.2$ for connected components and $\nu < -0.3$ for cavities. 
In the left panel, we observe that variations in the cosmological parameters generate an almost constant shift in $R_{\rm eff}^{{\rm con}}$ for $\nu > 1.4$. At lower thresholds $\nu \lesssim 1.4$, where the connected components develop more complex topology, the sensitivity of $R_{\rm eff}^{{\rm con}}$ to cosmological parameters becomes weaker. For cavities (cf. $R_{\rm eff}^{{\rm cav}}$, $\nu < 0$), the percolating regime does retain some cosmological parameter dependence. However, the blue, red and green curves all have similar shapes up to an overall amplitude and sign, indicating that this quantity also exhibits strong degeneracies between cosmological parameters. 

In the bottom middle and right panels, we show the parameter sensitivities of $\beta_{J}^{{\rm con}, {\rm cav} \, (1)}$ and $\beta_{J}^{{\rm con}, {\rm cav} \, (2)}$. The solid curves and shaded regions are the mean and error on the mean of $\beta_{1}^{{\rm con}, {\rm cav} \, (1)}$ and $\beta_{1}^{{\rm con}, {\rm cav} \, (2)}$, while the dashed lines (overlapping the solid lines) denote the means of $\beta_{2}^{{\rm con}, {\rm cav} \, (1)}$ and $\beta_{2}^{{\rm con}, {\rm cav} \, (2)}$. We find that both $\beta_{J}^{{\rm con}, {\rm cav} \, (1)}$, $\beta_{J}^{{\rm con}, {\rm cav} \, (2)}$ exhibit only very weak sensitivity to the cosmological parameters across all thresholds probed. This 
indicates that the average ellipticity of connected components and cavities is a characteristic property of Gaussian and weakly non-Gaussian fields smoothed with a Gaussian kernel, and is not strongly sensitive to the detailed shape of the underlying power spectrum. 

We quantify the degeneracy between different cosmological parameters in the sensitivity curves using a simple prescription. Although $N = 50$  realisations are insufficient to accurately construct the full covariance matrix, we can approximately explore degeneracies  using a Fisher-like inner product. As a first step, using $N=50$ realisations of the fiducial Quijote cosmological data, we evaluate the $Y_{k}$ statistic at $m=201$ equi-spaced $\nu_{i}$ threshold values, and define the diagonal $m\times m$ covariance matrix for the $Y_{k}$ statistic as  
\begin{equation} C^{(k)}_{ij} = {1 \over N-1} \sum_{n=1}^{N} [Y_{k, n}(\nu_{i}) - \bar{Y}_{k}(\nu_{i})]^{2} \delta_{ij},
\end{equation} 
where $Y_{k, n}(\nu_{i})$ is the value of the $Y_{k}$ statistic for the $n^{\rm th}$ realisation, in the $i^{\rm th}$ threshold bin and $\bar{Y}_{k}(\nu_{i})$ is the realisation mean in the $i^{\rm th}$ threshold bin. Here, $\delta_{ij}$ denotes the Kronecker delta function. This covariance matrix ignores the cross-correlations between different threshold bins (i.e. the off-diagonal terms). Then, using the finite-difference sensitivities given by eq.~(\ref{eq:sens}), we define a dimensionless, normalised Fisher-like  matrix
\begin{equation} \hat{F}^{(k)}_{\mu\nu} = { 1 \over A_{\mu\nu}} \bigg\langle {\partial Y_{k} \over \partial \theta_{\mu}} \bigg\rangle (C^{(k)})^{-1}  \bigg\langle{\partial Y_{k} \over  \partial \theta_{\nu}} \bigg\rangle, 
\end{equation} 
where $\mu, \nu$ run over the three cosmological parameters, $\langle.\rangle$ denote average over 50 realizations, and the normalisation factor is defined as
\begin{equation} A_{\mu\nu} = \sqrt{\left(  \bigg\langle {\partial Y_{k} \over \partial \theta_{\mu}} \bigg\rangle (C^{(k)})^{-1}  \bigg\langle {\partial Y_{k} \over \partial \theta_{\mu}} \bigg\rangle\right) \left(  \bigg\langle{\partial Y_{k} \over \partial \theta_{\nu}}\bigg\rangle (C^{(k)})^{-1}  \bigg\langle{\partial Y_{k} \over \partial \theta_{\nu}} \bigg\rangle\right)}.
\end{equation}

The eigenvalues of $\hat{F}^{(k)}$ quantify the degeneracies of the $Y_{k}$ statistic with respect to the three cosmological parameters, with correlations across thresholds neglected. While this simplification is too crude for accurate parameter estimation, it is sufficient for assessing the degeneracy structure encoded in $\hat{F}_{\mu\nu}^{(k)}$. If all three cosmological parameters have a strongly degenerate impact on $Y_{k}$, then the eigenvalues will take the form $(\lambda_{1},\lambda_{2},\lambda_{3}) \simeq (3, \epsilon, \epsilon')$ with $\epsilon, \epsilon' \ll 1$. In the opposite limit of no degeneracy, the eigenvalues would be $(\lambda_{1},\lambda_{2},\lambda_{3}) \simeq (1, 1,1)$.

\begin{table}[h!]
\centering
\begin{tabular}{| c | c c c c c c c c c c |}
\hline
$Y_{k}$   & $W_{0}$        & $W_{1}$    & $W_{2}$ & $W_{3}$ & $W_{1}^{0,2}|_{1}^{1}$ &  $W_{2}^{0,2}|_{1}^{1}$ & $b_{2}$ & $b_{0}$ & $R_{\rm eff}^{{\rm con}}$ &  $R_{\rm eff}^{{\rm cav}}$ \\ \hline
$\lambda_{1}$ & $3.0$   & $2.72$  & $2.49$ & $2.44$ & $2.77$ & $2.52$ & $2.92$ & $2.21$ & $2.94$ & $2.81$ \\
$\lambda_{2}$ & $10^{-3}$   & $0.28$  & $0.50$ & $0.55$ & $0.23$ & $0.48$ & $0.07$ & $0.78$ & $0.06$ & $0.17$ \\
$\lambda_{3}$ & $10^{-6}$   & $10^{-5}$  & $10^{-4}$ & $10^{-4}$ & $10^{-4}$ & $10^{-3}$ & $0.01$ & $10^{-3}$ & $10^{-3}$ & $0.01$ \\
\hline 
\end{tabular}
\caption{Eigenvalues $\lambda_{1}, \lambda_{2}, \lambda_{3}$ of the degeneracy matrix $\hat{F}^{(k)}_{\mu\nu}$ for each $Y_{k}$ statistic measured in this work. }
\label{tab:evals}
\end{table}

In Table~\ref{tab:evals} we present the three eigenvalues of $\hat{F}^{(k)}_{\mu\nu}$ for the statistics used in this work. We do not include the eigenvalue ratios $\beta_{J}^{{\rm con}, {\rm cav} \, (1,2)}$, as their sensitivity curves are practically consistent with zero and the corresponding $\hat{F}_{\mu\nu}^{(k)}$ matrices are numerically unstable.    
The results show that variations of $\Omega_{m}$, $n_{s}$ and $\sigma_{8}$ have an almost perfectly degenerate effect on $W_{0}$. The other volume-averaged statistics perform better, with two distinct eigenvalues $\lambda_{1} > \lambda_{2} \gg \lambda_{3}$. This indicates that one of the degeneracy directions is broken : $\sigma_{8}$ generates a distinct imprint on all $Y_{k}$ curves, except for $W_{0}$. However, the $\Omega_{m}$-$n_{s}$ degeneracy remains unbroken. The Betti number $b_{0}$ has a larger $\lambda_{2}$ and smaller $\lambda_{1}$ relative to $W_{3}$. This is consistent with the expectation that Betti numbers encode additional topological information, and hence more discriminatory power,  beyond that captured by the MFs. The Minkowski Functionals $W_{1-3}$ perform comparably, with $W_{3}$ exhibiting marginally higher sensitivity to $\sigma_{8}$. The Minkowski tensor $W_{1}{}^{0,2}$ and $W_{2}{}^{0,2}$ yield eigenvalues similar to their corresponding scalar functionals $W_{1}$ and $W_{2}$, as expected. 

The strong degeneracies within the $Y_{k}$ summary statistics suggest that constraints on cosmological parameters are not much improved even when the morphological and topological statistics are combined. However, they provide complementary information to the two-point statistics, and we expect that a combination of these measurements would reduce parameter uncertainties~\cite{Liu:2022vtr,Liu:2023qrj,Liu:2024uxa}. 

\section{Discussion} 
\label{sec:discuss}

\subsection{Summary of Results}
We have investigated three different cosmological effects : non-linear gravitational collapse, redshift space distortion (RSD) and massive neutrino free streaming, and studied their imprints on the morphology of three-dimensional matter density fields. Using  a set of global and local morphological descriptors including the Minkowski Functionals, Minkowski Tensors, Betti numbers, and the effective radius and shape of connected components and cavities, we have shown that each process imprints a characteristic and qualitatively distinct signature on excursion-set geometry and topology.

Gravitational collapse primarily induces non-Gaussianity through mode coupling, leading to skewness in the Minkowski Functional curves and significant modifications to the Betti numbers, while preserving statistical isotropy at the global level.  Massive neutrinos suppress small-scale power in an isotropic fashion, leading to subtle changes; primarily modifying one-point cumulants and decreasing the amplitude of the MFs. In contrast, redshift-space distortions strongly modify the shapes of individual excursion sets and directionally sensitive global measures such as the MTs. 

The percolation of the field from high to low thresholds imprints a particular pattern on the average size and shape of individual connected components and cavities (cf. bottom panels, Figure~\ref{fig:1_global}). The peaks at $\nu = \pm 1.4$ correspond to the threshold at which large excursion sets amalgamate, and is slightly lower than the locations of peaks of the Betti numbers $b_{0}$ and $b_{2}$. The position of the peaks, and in particular the shape of the $R_{\rm eff}^{{\rm con}}$ curve, is practically unaffected by gravitational collapse, redshift space distortion and the presence of massive neutrinos, after the one point skewness of the field has been removed. It is possible that the peak position could be estimated using an ellipsoid packing algorithm. The shape statistics $\beta_{J}^{{\rm con}, {\rm cav} \, (1,2)}$ are modified by gravitational collapse and redshift space distortion, in particular the connected components are both systematically more spherical in real space as a result of gravitational evolution while simultaneously being elongated marginally by redshift space distortion, with the largest effect at high thresholds. The shape of the cavities, as characterised by $\beta_{J}^{{\rm cav} \, (1,2)}$, are largely unchanged by the processes studied in this work, becoming marginally more anisotropic due to gravitational evolution. 

\subsection{Comparison with previous work}

Minkowski functionals have been widely used to quantify non-Gaussianity and non-linear structure formation in the matter and galaxy distributions, and have been shown to complement two-point statistics by capturing one-point information. More recently, Minkowski tensors have been developed and applied as a natural extension of scalar Minkowski functionals, providing direct sensitivity to directional information and anisotropy.

Our findings are consistent with earlier investigations of topology of LSS of the universe  but extend them in several directions.
A direct parallel can be drawn between our results and the earlier topological analysis of~\cite{2005ApJ...633....1P}, where authors examined how gravitational evolution, biasing, and redshift-space distortions modify the genus statistic. They found that gravitational nonlinearity significantly alters the topology by reducing the number of isolated high-density regions and enhancing void merging; trends fully consistent with our Betti-numbers and effective-radius results. Their conclusion that RSD lowers the genus amplitude without significantly altering its qualitative shape mirror our finding that RSD suppresses Minkowski Functionals and produces strong anisotropy captured by the Minkowski Tensors. Thus, our findings not only corroborate the trends identified in~\cite{2005ApJ...633....1P} but also extend them by incorporating an expanded set of morphological statistics. Our RSD analysis builds on earlier work demonstrating the ability of Minkowski Tensors to capture anisotropy induced by coherent flows and virial motions due to peculiar velocities~\cite{2018ApJ...863..200A,Appleby:2022itn}. Our results show that redshift space introduces a measurable anisotropy in the Minkowski Tensors, with the line-of-sight component deviating significantly from the transverse components, breaking the degeneracy seen in real space. 

The morphological impact of massive neutrinos has been explored in a limited number of previous works, primarily using Minkowski Functionals applied to weak-lensing fields \cite{PhysRevD.99.083508,Marques:2018ctl} or three-dimensional matter and galaxy density fields \cite{Liu:2022vtr,Liu:2023qrj}. These studies focus on parameter inference and demonstrate that combining morphological statistics with the power spectrum or bispectrum can significantly improve constraints on the sum of neutrino masses. In contrast, the focus of our analysis is on developing a qualitative understanding of how different cosmological effects modify the excursion-set morphology. Within this framework, we find that massive neutrinos leave the Minkowski Tensors isotropic, introduce only mild but systematic changes in the MFs and Betti numbers, primarily due to the suppression of small scale power. A multi-scale probe of the density field would allow us to better study the impact of neutrino mass and will be considered elsewhere.

\subsection{Sensitivity of Morphological Statistics to Cosmology}

Our analysis shows that the global morphological statistics, particularly the amplitudes of the Minkowski Functionals, are sensitive to variations in cosmological parameters such as $\Omega_m$, $n_s$ and $\sigma_8$. While $\Omega_m$ and $n_s$  modify the shape of the matter power spectrum, $\sigma_8$ rescales the overall amplitude of density fluctuations. These changes lead to coherent modifications in the geometry and topology of excursion sets. A detailed cosmological parameter analysis exploiting these sensitivities will be presented in a forthcoming study.

In contrast, the local statistics like effective radius $R^{{\rm con}, {\rm cav}}_{\rm eff}$ and shape parameters $\beta_{J}^{{\rm con}, {\rm cav} \, (1,2)}$ exhibit very limited sensitivity to cosmological parameter variations. However, these same local descriptors are affected by the physical effects examined here: gravitational collapse produces marked asymmetries between overdense and underdense regions, while RSDs generate measurable anisotropic distortions of size and shape of structures. Their sensitivity to such physical processes, despite their insensitivity to cosmological parameters, indicates that they probe complementary aspects of structure formation that are not captured by global, averaged morphological measures. 

Regarding the cosmological parameter sensitivity of the MFs, we note that our sensitivity curves $\partial Y_{k}/\partial\theta$ differ in shape compared to other work in the literature \cite{Liu:2022vtr,Liu:2023qrj}. Those works use a different threshold $\nu = \rho/\bar{\rho} = 1 + \delta$ to define excursion sets, where $\bar{\rho}$ is the mean density of the field. This should be compared to our choice $\nu = \delta/\sigma_{0}$. The difference is not simply a common re-scaling of the $x$-axis for non-Gaussian fields, and will change the structure of the curves as we vary the cosmological parameters. Although any reasonable choice of excursion set thresholding can be used, and in fact the one adopted in \cite{Liu:2022vtr,Liu:2023qrj} may help to break certain parameter degeneracies, we choose the definition for which theoretical results are typically calculated \cite{2003ApJ...584....1M}.

\subsection{Limitations \& Future work}
The present analysis is restricted to dark-matter-only simulations and idealized density fields. Extensions to biased tracers, realistic survey geometries and shot noise will be essential for direct application to observational data. In addition, while we have explored the response to a subset of cosmological parameters, future work will focus on joint inference using full covariance information and likelihood-based or machine-learning-driven approaches.

\acknowledgements
{PG is supported by KIAS Individual Grant (PG088101) at Korea Institute for Advanced Study (KIAS). PC gratefully acknowledges the support of a visiting professorship at KIAS, during which part of this work was carried out. CBP is supported by the
National Research Foundation of Korea (NRF) grant funded by
the Korean government (MSIT; RS-2024-00360385).}

\appendix 

\section{Effect of Gravitational Evolution}
\label{app:a}

As stated in Section~\ref{sec:grav}, the overwhelming effect of gravitational collapse is to skew the density threshold $\nu$, as seen for example in the top-left panel of Figure~\ref{fig:2_global} showing $W_0$, whose shape traces the cumulative probability distribution function (CPDF).  Gravitational collapse changes the one-point function of the density field from a Gaussian form to one that is highly skewed towards large positive values. The non-Gaussianity in the one-point function is eliminated in Section~\ref{sec:grav} via a $\nu \to \nu_{v}$ threshold transformation. 

\begin{figure}[h!]
    \centering
    \includegraphics[width=0.5\textwidth]{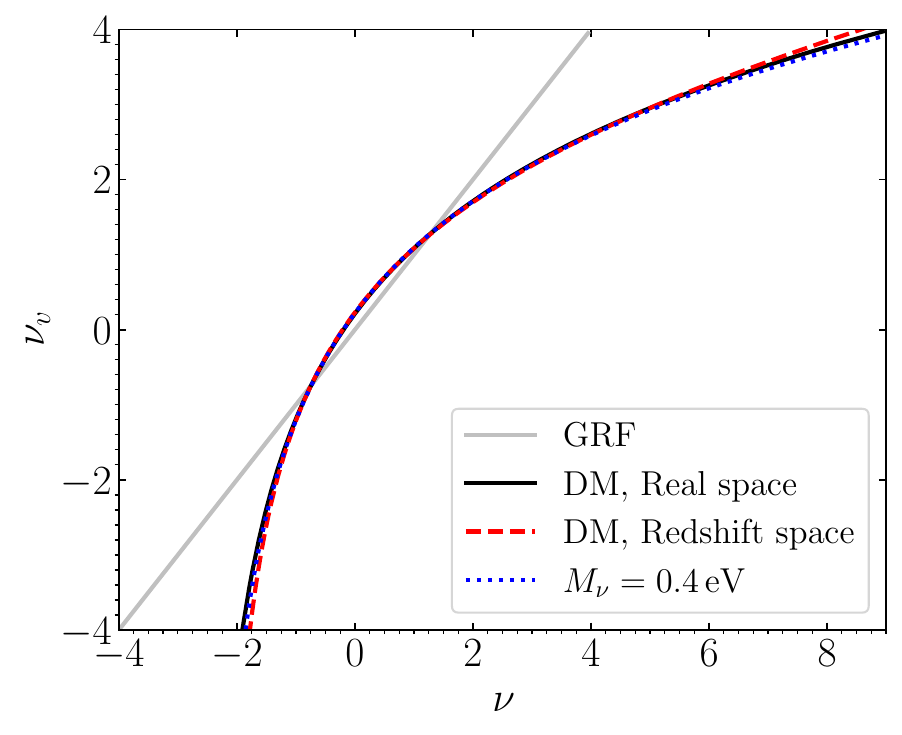}
    \caption{Volume fraction threshold $\nu_{v}$ presented as a function of iso-field threshold $\nu$ for the $z=0$, real space dark matter fields (black curve) and Gaussian random fields (grey curve).}
    \label{fig:nu_nuv}
\end{figure}

In Figure~\ref{fig:nu_nuv}, we present $\nu_{v}$ as a function of $\nu$ for the $z=0$ dark matter density field in real space (solid black curve), in redshift space (red dashed curve), and in real space with a massive neutrino component (blue dotted curve). These curves fully characterize the evolution of the density field PDFs from Gaussian to skewed distributions. The Gaussian expectation, $\nu_{v} = \nu$, is shown by the grey line for reference.

\section{Information Content of the Local Statistics}
\label{sec:app_c}

Here we briefly outline how ensemble averages of the Minkowski functionals and tensors differ from averages over the morphology of individual connected components or cavities.  We focus on $R_{\rm eff}^{{\rm con}}$ and $W_{0}$, as they are the simplest statistics that can be used to demonstrate this distinction. The key difference arises because, although both $R_{\rm eff}^{{\rm con}}$ and $W_{0}$ can be expressed as averages over excursion sets, only $W_{0}$ can additionally be written as a volume average over pixels. Consequently, for $W_{0}$, ensemble and volume averages commute, whereas for $R_{\rm eff}^{{\rm con}}$ they do not. 

For simplicity, we consider Gaussian random fields here. Let us recall the definition of $R_{\rm eff}^{{\rm con}}$:
\begin{equation}
     R_{\rm eff}^{{\rm con}} = {1 \over b_{0}}\left( {3 \over 4\pi}\right)^{1/3} \sum_{i=1}^{b_{0}} x_{i} \quad ,
\end{equation}
where $b_{0}$ is the number of connected components, and $x_i$ is the cube root of the volume $V_i$ of the $i$th connected component $Q_i$,
\begin{equation} x_{i} = V_{i}^{1/3} = \left( \int_{Q_{i}}dV\right)^{1/3} \, .
\end{equation} 
Note that $R_{\rm eff}^{{\rm con}}$ is a function of the random variables $b_{0}$ and $x_i$. We also recall that $W_{0}$ is given by: 
\begin{equation}W_{0} = {1 \over V} \sum_{i=1}^{b_{0}} V_{i} \  , 
\end{equation} 
and is therefore a function of the random variables $V_i$.

The ensemble averages of $R_{\rm eff}^{{\rm con}}$ and $W_{0}$ are given by 
\begin{eqnarray} 
\label{eq:reff_th}  \langle R_{\rm eff}^{{\rm con}} \rangle &=& \left( {3 \over 4\pi}\right)^{1/3} \left\langle {1 \over b_{0}} \sum_{i=1}^{b_{0}} x_{i} \right\rangle = \left( {3 \over 4\pi}\right)^{1/3} \sum_{n\geq 1} P(b_{0}=n \, | \, V) \left\langle {1 \over n} \sum_{i=1}^{n} x_{i} \bigg| b_{0}=n \, , \, V \right\rangle  \\ 
\label{eq:f1} &=& \left( {3 \over 4\pi}\right)^{1/3} \sum_{n\geq 1} P(b_{0}=n \, | \, V) \left\langle  x_{1} \big| b_{0}=n \, , \, V \right\rangle \, , \\ 
\label{eq:appc_1}  \langle W_{0} \rangle &=& {1 \over V} \left\langle  \sum_{i=1}^{b_{0}} V_{i} \right\rangle = {1 \over V} \sum_{n\geq 1} P(b_{0}=n | V) \left\langle  \sum_{i=1}^{n} V_{i} \bigg| b_{0}=n, \, V \right\rangle  \\
\label{eq:f2} &=& {1 \over V} \sum_{n\geq 1} P(b_{0}=n | V) \, n \, \left\langle  V_{1} \big| b_{0}=n, \, V \right\rangle \, . 
\end{eqnarray} 
\noindent In the above, we have decomposed the ensemble averages into contributions conditional on $b_{0}$ while keeping the total volume $V$ finite. The quantity $P(b_{0}=n | V)$ denotes the probability of $n$ connected components contained within a volume $V$. In eqs. (\ref{eq:f1},  \ref{eq:f2}) we have commuted the sums over $n$ and the ensemble averages, and then reduced the sums to $n$-copies of $\left\langle  x_{1} \big| b_{0}=n, \, V \right\rangle$ and $\left\langle  V_{1} \big| b_{0}=n, \, V \right\rangle$. This last step can be taken because the joint distribution of $\{x_{i}\}$ (or $\{V_{i}\}$), conditioned on $b_{0}=n$, is invariant under permutations of the index $i$, which allows the sum to be reduced to $n$ identical conditional expectations. 

For $W_{0}$, we can re-write the sum over excursion sets as the sum over pixels as
\begin{equation} \sum_{i=1}^{b_{0}} V_{i} = \sum_{i=1}^{N_{\rm pix}} \Delta^{3} \Theta(\delta_{i} - \nu) \,
\end{equation}
where $\Theta(x)$ is the Heaviside function and $\Delta^{3}$ is the pixel size. With this re-writing, the only random variables after the first equality of eq.~(\ref{eq:appc_1}) are the pixel values $\delta_{i}$ in the Heaviside function (and specifically the sum is now over the deterministic quantity $N_{\rm pix}$), so the ensemble average commutes with the sum and we can write
\begin{equation}\label{eq:w0} \langle W_{0} \rangle = {\Delta^{3} \over V} \sum_{i=1}^{N_{\rm pix}} \int_{-\infty}^{\infty} P(\delta_{i}) \Theta(\delta_{i} - \nu) {\mathrm{d}}\delta_{i} \, , \end{equation} 
from which the standard result follows -- the sum is simply $N_{\rm pix}$ copies of the one point integral of the density field PDF integrated over the range $(\nu,\infty)$, and we can take the limits $N_{\rm pix} \to \infty$, $V = N_{\rm pix} \Delta^{3} \to \infty$. 

The reason why we can exactly calculate $\langle W_{0} \rangle$ is the linearity of the sum, and the fact that we can re-write the sum over excursion set volumes as a sum over all pixels. This is because $W_{0}$ contains no information about correlations between pixel values, it depends  only on whether each pixel is `in' or `out' of the excursion set. 

Although the expression (\ref{eq:f1}) is exact for $\langle R_{\rm eff}^{{\rm con}}\rangle$, it cannot be reduced further to a simple closed form like $\langle W_{0}\rangle$ in eq.~(\ref{eq:w0}). If we assume that $b_{0}$ and $V_{i}$ are independent variables, and $V_{i}$ are independent and identically distributed (iid), then we could proceed by writing (\ref{eq:f1}) as : 
\begin{equation}\label{eq:reff_approx}  \langle R_{\rm eff}^{{\rm con}} \rangle   =  \left( {3 \over 4\pi}\right)^{1/3} \left\langle x_{1} \, | \, V   \right\rangle \, .
\end{equation} 
That is, under these assumptions $\langle R_{\rm eff}^{{\rm con}} \rangle$ would be a measure of a fractional moment $\langle V_{1}^{1/3}\rangle$ of the probability density function $P(V_{1})$ of connected component volumes. However, the assumptions used to arrive at eq.~(\ref{eq:reff_approx}) are not reliable over the entire density threshold range that we consider. 

We examine these assumptions in turn. First, the number of  connected components $b_{0}$ is correlated with the volume of individual elements $V_{i}$ for all thresholds, and the correlation is particularly strong for $0 \lesssim \nu \lesssim 1.4$. At large thresholds $\nu > 2$, the correlation is weak and due to the constraint that the total volume of all connected components  must equal $W_{0}$, so increasing $b_{0}$ forces the average volume of subsets to decrease. For low thresholds $0 \lesssim \nu \lesssim 1.4$, the excursion set is dominated by a small number of large-volume subsets together with many smaller disconnected regions. In this regime, the random variable $b_{0}$ is strongly influenced by the presence of these large structures, which start to form at $\nu \lesssim 1.4$. 

Second, for defining the ensemble average of the Minkowski Functionals, we take the limit $V \to \infty$. In terms of the ensemble average over $b_{0}$ (cf. eq.~(\ref{eq:f1})), this would ideally correspond to the limit $\langle b_{0} \rangle \propto V \to \infty$. In practice, we estimate summary statistics from finite-volume fields, and in the limits $\nu \gg 1$ and $\nu \lesssim 0$, we will measure only a small number of subsets. In particular, for $\nu \lesssim 1.4$, the sub-volumes merge into a small number of (or single) percolating structures, and it is unclear how well volume and ensemble averages will match. Similarly, for finite $V$ we will always encounter a large threshold limit where the number of excursion sets is small. For both limits, we should compare measured averages from data with eq.~(\ref{eq:f1}), which makes no assumption about the size of $V$. 

Finally, we turn to the assumption that $x_{i}$ are iid. This is likely to be a good approximation for high thresholds where  connected components  are spatially well separated. In contrast, at $\nu \lesssim 1.4$, large volume subsets form and percolation occurs, in which case the distribution of $x_{i}$ will be characterized by a small number of large volume connected components and more numerous, small volume islands. The large and small volume elements are neither independent nor identically distributed, making an estimate of $\langle x \rangle$ difficult. 

\begin{figure}[h!]
    \centering
    \includegraphics[width=0.98\textwidth]{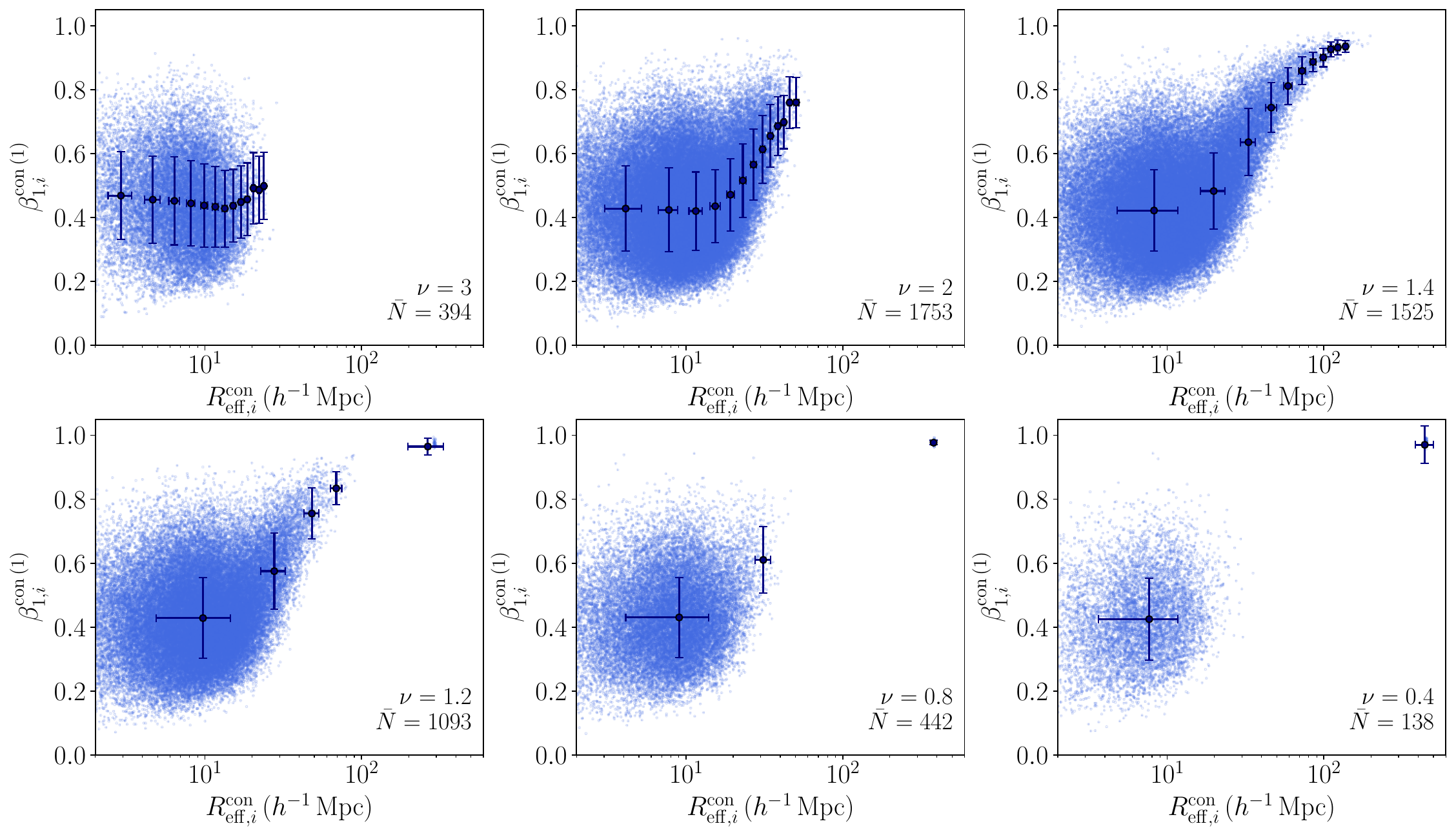}
    \caption{Two-dimensional distribution of $\beta_{1, i}^{{\rm con} \, (1)}$ and $R_{{\rm eff}, i}^{{\rm con}}$ extracted from individual connected components from $N_{\rm real} = 50$ realisations of a Gaussian random field in real space (blue points). Each blue point represents a single excursion region from the realisations. The top row (left-right panels) are  connected components at thresholds $\nu = 3, 2, 1.4$ and the bottom row $\nu = 1.2, 0.8, 0.4$. The dark blue points/errorbars are the mean and $1\sigma$ uncertainty of the binned point cloud.}
    \label{fig:app_1}
\end{figure}

To highlight these issues, in Figure~\ref{fig:app_1} we present the scatter plots of the effective radius $R_{\rm eff}^{{\rm con}}$ and the shape parameter $\beta_{1}^{{\rm con} \, (1)}$ for all connected components measured from $N_{\rm real}=50$ realizations of real space Gaussian random fields using the same parameters and smoothing as in Section~\ref{sec:grf}. Each blue point corresponds to a single  connected component, and each panel shows a different density threshold (top row, left to right: $\nu=3,2,1.4$; bottom row, left to right: $\nu=1.2,0.8,0.4$). The dark blue points with error bars show the binned mean and $1\sigma$ scatter. We display $\beta_{1}^{{\rm con} \, (1)}$ as representative of the four shape parameters measured in this work; the remaining three exhibit qualitatively similar behaviour.

Starting from a high threshold $\nu=3$ (top left panel), the  connected component are drawn from a relatively compact two-dimensional distribution with only weak correlation between $R_{\rm eff}^{{\rm con}}$ and $\beta_{1}^{{\rm con} \, (1)}$\footnote{A small systematic trend of $\beta_{1}^{{\rm con} \, (1)}$ first decreasing then increasing with increasing $R_{\rm eff}^{{\rm con}}$ is visible in the binned averages.}. These objects are predominantly isolated peaks of the Gaussian field and are approximately independent realisations of a single population.

As the threshold is lowered, this simple picture breaks down. By $\nu=1.4$ (top right panel) a second population of connected components emerges, characterised by substantially larger radii and higher values of $\beta_{1, i}^{{\rm con} \, (1)}$. These objects are no longer simple peak-like structures but are composite regions formed from chains of peaks connected by saddle critical points. The effective sizes of connected components are,  therefore, no longer drawn from a single underlying distribution, and the assumption that they form an approximately iid ensemble becomes invalid.

Upon further decreasing the threshold (bottom row), a single percolating object appears in each realisation,  visible as isolated points in the top-right corner of each panel. As $\nu$ decreases from $1.2$ to $0.8$ and $0.4$ (left to right panels), this dominant object grows by absorbing an increasing fraction of the remaining smaller subsets. Hence for $\nu\lesssim 1.2$, the excursion set bifurcates into one extremely large region and an increasingly sparse population of smaller objects. The global Minkowski functionals $W_{0}$--$W_{3}$, being integrals over the entire excursion set volume/boundary, become dominated by the properties of the single percolating object at $\nu \lesssim 1.2$, whereas local statistics such as the shape parameters $\beta_{J, i}^{{\rm con} \, (1,2)}$ probe predominantly the morphology of the small, peak--like point clouds. 

The scatter plots shown in Figure~\ref{fig:app_1} make it explicit that below the percolation threshold the  connected components cannot be treated as a homogeneous ensemble, which obfuscates the analytic modelling of their average morphological properties. However, one could introduce cuts, in $R_{\rm eff}^{{\rm con}}$ for example, to isolate connected components that are approximately iid and hence more amenable to analytic study.

\section{High Threshold Ensemble Average}
\label{sec:app_th}

We can use the local, elliptical peak approximation to estimate the expectation value of $\langle R_{\rm eff}^{{\rm con}} \rangle$, $\langle \beta_{J}^{{\rm con} \, (1)}\rangle$, and $\langle \beta_{J}^{{\rm con} \, (2)}\rangle$ at high thresholds. In this Appendix, we restrict our analysis to Gaussian random fields in real space, for which the statistical properties of peaks are well studied \cite{Bardeen:1985tr,Codis:2013exa,Germani:2025fkh}. 

A peak (located at some arbitrary point ${\bf r}= {\bf 0}$) can be characterized by the field $\delta$, and its first $\eta_{i} = \nabla_{i}\delta$ and second $\xi_{ij} = \nabla_{i}\nabla_{j} \delta$ derivatives at that point. In particular, a peak is defined by $\eta_{i}({\bf 0}) = 0$ and $\xi_{ij}$ having three negative eigenvalues -- these correspond to conditions on the joint ten-dimensional PDF $P(\delta, \eta_{i}, \xi_{ij})$ that can be partially marginalised over. We can write a local expansion of the density in the vicinity of a peak of height $\delta_{\rm pk}$ at ${\bf r} = {\bf 0}$ : 

\begin{equation}
 \label{eq:d_exp}   \delta({\bf r}) = \delta_{\rm pk} + {1 \over 2} \sum_{i  , j =1}^{3} \xi_{ij} r^{i} r^{j} + {\cal O}(|{\bf r}|^{3})
\end{equation}
It is convenient to define all variables in dimensionless form as $\nu = \delta/\sigma_{0}$, $\tilde{\eta}_{i} = \eta_{i}/\sigma_{1}$ and $\tilde{\xi}_{ij} = -\xi_{ij}/\sigma_{2}$, where 
\begin{equation} \sigma_{i}^{2} = {1 \over 2\pi^{2}} \int k^{2+2i} dk P(k) W(k R_{\rm G})^{2} \, ,
\end{equation}
and we also define $\gamma = \sigma_{1}^{2}/(\sigma_{0}\sigma_{2})$. We have used the standard convention of introducing a minus sign in the definition of $\tilde{\xi}_{ij}$ so that the eigenvalues for a peak are positive. Next we rotate the local coordinate system such that $\tilde{\xi}_{ij}$ is diagonal, with components corresponding to its eigenvalues $\lambda_{1} \geq  \lambda_{2} \geq \lambda_{3} > 0$, and define the following variables 

\begin{equation} x = \lambda_{1} + \lambda_{2} + \lambda_{3} \, , \, \quad e = {\lambda_{1} - \lambda_{3} \over 2x} \, , \, \quad p = {\lambda_{1} - 2\lambda_{2} + \lambda_{3} \over 2 x} \, , \,
\end{equation} 
In terms of these variables, the peak probability distribution of height $\nu_{\rm pk}$, and shape parameters $x$, $e$, $p$ is given by \cite{Bardeen:1985tr,Germani:2025fkh} 
\begin{equation} P_{\rm pk} (\nu_{\rm pk}, x, e, p) = A x^{8} e^{-Q_{\nu x}(\nu_{\rm pk}, x)} e^{-Q_{e p}(x, e,p)} {\cal J}(e,p) \Theta(x,e,p) ,
\end{equation} 
where 
\begin{equation} Q_{\nu x}(\nu_{\rm pk}, x) = {\nu_{\rm pk}^{2} - 2\gamma \nu_{\rm pk} x + x^{2} \over 2 (1 - \gamma^{2})} \, , \, \qquad \, Q_{ep}(x,e,p) = {5 \over 2} x^{2} (3e^{2}+ p^{2})  \, , 
\end{equation} 
and 
\begin{equation}
    {\cal J}(e,p) = e (e^{2} - p^{2}) (1-2p) \left[(1+p)^{2} - 9 e^{2} \right],
\end{equation}
and 
\begin{equation} \Theta(x,e,p) = \begin{cases} 
    1  & {\rm if\ } (0 \leq e \leq 1/4\, , \quad -e \leq p \leq e \, , \quad x > 0), \\ 
    1  & {\rm if\ }  (1/4 \leq e \leq 1/2\, \quad 3e-1 \leq p \leq e \, , \quad x > 0), \\ 
    0  & {\rm otherwise}. 
\end{cases}
\end{equation} 
The normalisation factor, $A$, is defined such that the PDF integrates to unity, but cancels for the conditional probabilities that we will consider below. 

For a given threshold $\delta({\bf r}) = \nu \sigma_{0}$, we can re-write equation (\ref{eq:d_exp}) as 
\begin{equation} \nu = \nu_{\rm pk} - {\sigma_{2} \over 2\sigma_{0}} \sum_{i} \lambda_{i}q_{i}^{2} , \end{equation} 
with $\nu_{\rm pk} > \nu$ and $q_{i}$ are the rotated coordinates aligned with the ellipsoid principal axes. The volume of the ellipsoid enclosed by $\nu$ can be inferred from this equation in terms of the variables $\nu$, $\nu_{\rm pk}$, $x$, $e$, $p$ : 
\begin{equation} V \simeq {4 \pi \over 3} \left({2 \sigma_{0} \over \sigma_{2}}\right)^{3/2} (\nu_{\rm pk} - \nu)^{3/2} \left({3  \over x}\right)^{3/2} {1 \over \sqrt{(1 + 3e + p)(1 - 2p)(1-3e+p)}}.
\end{equation} 
We can approximate the expectation value for $\langle R_{\rm eff}^{{\rm con}} \rangle$ as
\begin{eqnarray}
 \label{eq:reff}   \langle R_{\rm eff}^{{\rm con}} \rangle &\simeq& {1 \over N} \int_{\nu}^{\infty} \,  \d\nu' \int \d x \, \d e \, \d p \, P_{\rm pk}(\nu', x, e, p) V^{1/3},  \\ 
   N &=&  \int_{\nu}^{\infty} \,  \d\nu' \int \, \d x \, \d e \, \d p P_{\rm pk}(\nu', x, e, p).
\end{eqnarray}
If the majority of the peaks lie close to $\nu$, and the number of filamentary saddle points with $\nu_{\rm sad} > \nu$ is much smaller than the number of peaks with $\nu_{\rm pk} > \nu$, then we can expect eq.~(\ref{eq:reff}) to provide a good approximation to the ensemble average. At thresholds for which the topology of excursion sets is non-trivial (a significant number of loops, for example) then this approximation will completely break down. 

\subsection{Local Shape Information}

We can perform a similar calculation for the $\beta_{J}^{{\rm con} \, (1)}$, $\beta_{J}^{{\rm con} \, (2)}$ shape functions. Parameterizing the coordinates on the surface of an ellipsoid as $x = a \sin \theta \cos \phi$, $y = b\sin\theta \sin \phi$, $z = c \cos\theta$ with $0 \leq \theta < \pi$ and $0 \leq \phi < 2\pi$ and $a \leq b \leq c$, the unit normal components to the surface are obtained by taking the gradient of the implicit equation of the ellipse :    
\begin{eqnarray} & &\hat{n}_{x} = {\sin\theta \cos\phi \over \ell} \, , \quad \hat{n}_{y} = {a \over b}{\sin\theta \sin\phi \over \ell} \, , \quad 
 \hat{n}_{z} ={a \over c} {\cos\theta \over \ell} ,\\
& & \ell(\theta,\phi) = \left[\left({a \over c}\right)^{2} \cos^{2}\theta +  \sin^{2}\theta \left(\cos^{2}\phi + \left({a \over b} \right)^{2} \sin^{2} \phi \right) \right]^{1/2},
\end{eqnarray} 
and the mean curvature of the surface is 
\begin{eqnarray}   G_{2} &=& {1 \over 2} \nabla . \hat{n} \\ 
 &=& {1 \over 2 \, a} \left[ {1 \over \ell(\theta,\phi)} \left( 1 + \left({a \over b}\right)^{2} + \left({a \over c}\right)^{2}\right) - {1 \over \ell(\theta,\phi)^{3}} \left( \left({a \over c}\right)^{4} \cos^{2}\theta  + \sin^{2}\theta \left( \cos^{2}\phi + \left({a \over b}\right)^{4} \sin^{2}\phi \right) \right) \right]. 
\end{eqnarray} 

Then, in this coordinate system the Minkowski Tensors are diagonal with components 

\begin{eqnarray}
     W_{1}^{0,2}|_{ii} &=& b c  \int_{0}^{2\pi} \d\phi \int_{0}^{\pi}\d\theta \sin\theta \, \ell(\theta, \phi)  \, \hat{n}^{2}_{i}, \\
     W_{2}^{0,2}|_{ii} &=& b c   \int_{0}^{2\pi} \d\phi \int_{0}^{\pi}\d\theta \sin\theta  \, \ell(\theta, \phi)  \, G_{2} \, \hat{n}^{2}_{i}.
\end{eqnarray}
\noindent For the ellipsoidal connected components described in the previous subsection, we can write the axis constants $(a,b,c)$ in terms of the random variables $\nu_{\rm pk}, x, e, p$ according to 
\begin{eqnarray}
    & & a^{2} = {6 (\nu_{\rm pk} - \nu) \over x(1+3e + p)} \, , \quad b^{2} = {6 (\nu_{\rm pk} - \nu) \over x(1 - 2 p)} \, , \quad c^{2} = {6 (\nu_{\rm pk} - \nu) \over x(1 - 3e + p)} \, .
\end{eqnarray}
\noindent The shape statistics, for a single excursion subset are $\beta_{J}^{{\rm con} \, (1)} = W_{1}^{0,2}|_{JJ}/W_{1}^{0,2}|_{x x}$, and $\beta_{J}^{{\rm con} \, (2)} = W_{2}^{0,2}|_{JJ}/W_{2}^{0,2}|_{x x}$.
The ensemble averages of the shape variables $\beta_{J}^{{\rm con} \, (1, 2)}$ are :
\begin{eqnarray} & &  \langle \beta_{J}^{{\rm con} \, (1, 2)} \rangle = {1 \over N} \int \d e \,\d p \, \beta_{J}^{{\rm con} \, (1, 2)}(e, p) {\cal J}(e,p) \int_{0}^{\infty} \d x \,\Theta(x, e, p)\, e^{-Q_{ep}(x, e, p)} x^{8} \int_{\nu}^{\infty}\d\nu' e^{-Q_{\nu x}(\nu', x)}, \\  
\nonumber & & N =  \int \d e\, \d p \,  {\cal J}(e,p) \int_{0}^{\infty} \d x \,\Theta(x, e, p) \,e^{-Q_{ep}(x, e, p)} x^{8} \int_{\nu}^{\infty} \d\nu' e^{-Q_{\nu x}(\nu', x)} .
\end{eqnarray} 
 \begin{figure}[h!]
    \centering
    \includegraphics[width=0.98\textwidth]{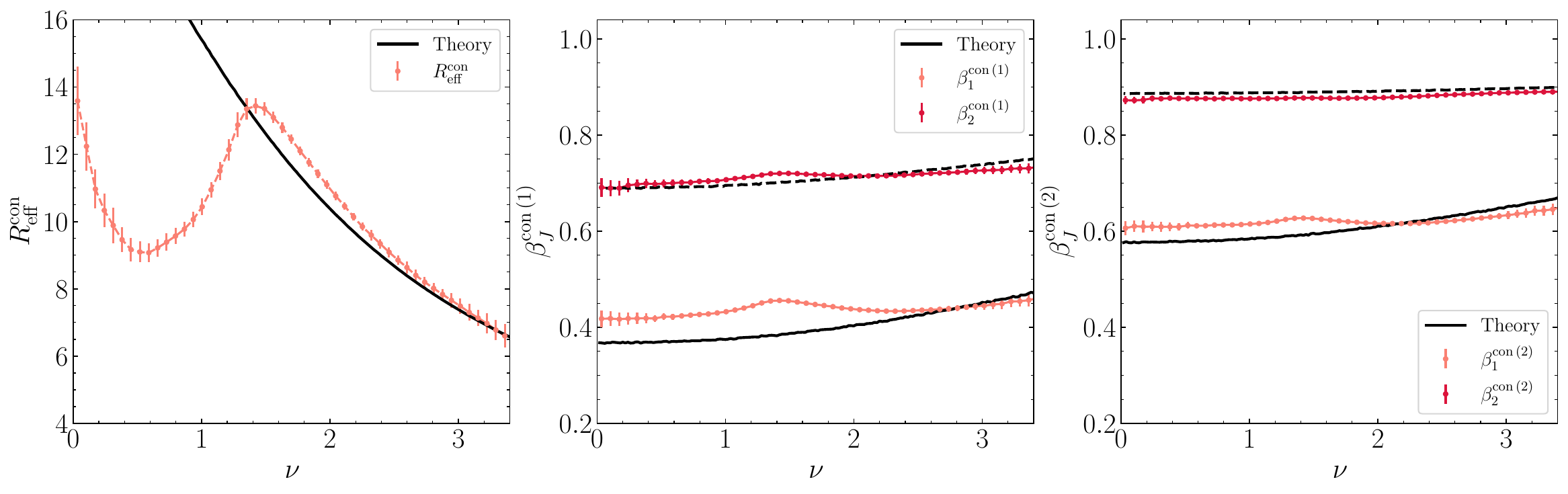}
    \caption{Comparison between measurements of $R_{\rm eff}^{{\rm con}}$, $\beta_{J}^{{\rm con}, {\rm cav} \, (1)}$ and $\beta_{J}^{{\rm con}, {\rm cav} \, (2)}$ (red points/error bars) extracted from $N=50$ Gaussian random fields and the theoretical predictions derived in Appendix~\ref{sec:app_th} (black solid curves).}
    \label{fig:th}
\end{figure}
All components of $\beta_{J}^{{\rm con} \, (1,2)}$ can be written purely in terms of axis ratios $a/b$ and $a/c$, and hence do not contain explicit $x$, $\nu_{\rm pk}$ or $\nu$ dependence. However, $\nu$ and $x$ still enter the ensemble averages due to the fact that they are correlated via the $\exp[-Q_{\nu x}]$ term, and $x$, $e$ and $p$ are subsequently correlated through $\exp[-Q_{ep}(x,e,p)]$ after $\nu$ has been marginalised over. 

In Figure~\ref{fig:th} we present measurements of $R_{\rm eff}^{{\rm con}}$ (left panel), $\beta_{J}^{{\rm con} \, (1)}$ (middle panel) and $\beta_{J}^{{\rm con} \, (2)}$ (right panel) for the Gaussian random fields considered in Section~\ref{sec:grf} (cf. red curves, lower panels of Figure~\ref{fig:1_global}). The black solid/dashed curves are the theoretical ensemble averages constructed in this section. We find good agreement for $\nu \gtrsim 2.5$, as expected. The numerical measurements of $R_{\rm eff}^{{\rm con}}$ (cf. red points, left panel) rise faster than the theoretical prediction, due to the presence of peak-saddle chains that have not been accounted for in our simple analytic prescription. 

The theoretical prediction does not exactly match $\beta_{J}^{{\rm con} \, (1)}$ or $\beta_{J}^{{\rm con} \, (2)}$ at high thresholds, with the black solid/dashed curves lying above the numerical results. This is due to a numerical systematic, which slightly reduces the isotropy of the connected components. Small, pixel-scale objects will be present at all thresholds, and these objects will exhibit spurious anisotropy due to the fact that they are not well resolved. This systematic presents a larger impact on $\beta_{J}^{{\rm con} \, (1)}$ and $\beta_{J}^{{\rm con} \, (2)}$ than dimension-full quantities such as $R_{\rm eff}^{{\rm con}}$, because barely resolved objects will be naturally down-weighted by their size in the $R_{\rm eff}^{{\rm con}}$ average. One could apply minimum size cuts to mitigate this numerical artifact. 

\begin{figure}[h!]
    \centering
    \includegraphics[width=0.5\textwidth]{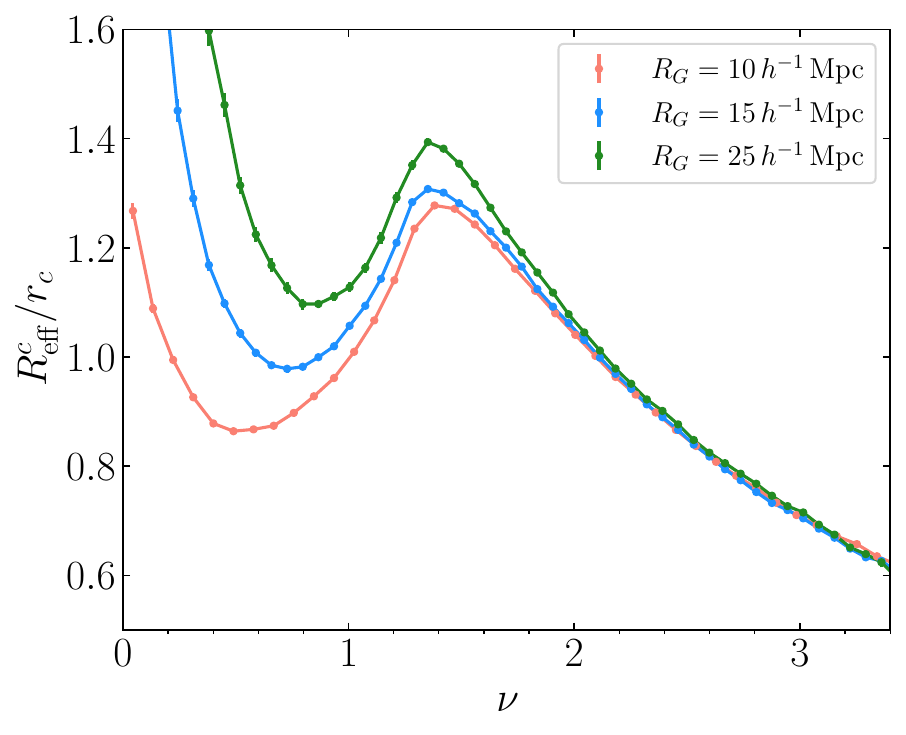}
    \caption{Measurements of $R^{{\rm con}}_{\rm eff}/r_{c}$ for $N=50$ realisations of Gaussian random fields, smoothed on scales $R_{G} = 10, 15, 25 \, h^{-1} \, {\rm Mpc}$ (red/blue/green points/errorbars).}
    \label{fig:appc_Reff}
\end{figure}

So far in this appendix, we have used an ellipsoidal approximation in the vicinity of a peak to estimate the morphological properties of  connected components. In Figure~\ref{fig:appc_Reff} we present an alternative perspective, specifically we plot the dimensionless quantity $R_{\rm eff}^{{\rm con}}/r_{c}$ over the range $0 \leq \nu \leq 3.5$ for a Gaussian random field in real space with the fiducial $\Lambda$CDM power spectrum from the main body of the text, where $r_{c} = \sigma_{0}/\sigma_{1}$. The red/blue/green points/error bars are the mean/error on mean of $N_{\rm real} = 50$ realisations of random fields smoothed on scales $R_{G} = 10, 15, 25 \, h^{-1} \, {\rm Mpc}$ respectively. We find that for $\nu \gtrsim 2$, the curves are practically indistinguishable. This indicates that to high precision, the statistic $R_{\rm eff}^{{\rm con}}$ can be described purely in terms of the correlation length of the field, $r_{c}$ and some function of $\nu$ at these thresholds.

Contrarily, at $\nu \lesssim 2$, $R_{\rm eff}^{{\rm con}}/r_{c}$ shows dependence on the smoothing scale $R_{G}$, with smaller scale smoothing dropping lower before finally rising at $\nu \sim 0$. This indicates that a higher number of small sized connected components survive after the large objects form at the peak $\nu \sim 1.4$ when we smooth with a smaller scale. In this regime $0 < \nu \lesssim 1.4$, the shape of the curve contains information on the percolation behaviour of the field. For $\nu \sim 0$, the rising behaviour common to all three curves is due to the smaller substructures combining with the single large excursion subset. 

In terms of the ensemble average discussion in this section, for $\nu \gtrsim 2$ the similarity of the curves suggests that the average morphology of the excursion sets is dominated by the local shape of the field encoded in $r_{c}$. This implies that we can approximately express $\langle R_{\rm eff}^{{\rm con}}\rangle$ purely in terms of an integral of the joint PDF $P(\delta,\nabla\delta, \nabla\nabla\delta)$ at a single point in the space, similarly to the Minkowski Functionals. The curves start to deviate from one another at $\nu \lesssim 2$, approximately where loops start to significantly contribute to the genus in the form of the Betti number $b_{1}$. 

\bibliography{cleaned_merged_SAFE}{}
\bibliographystyle{JHEP}

\end{document}